\documentclass[ 
	aps,
	footinbib,
	pra,
	twocolumn,
	superscriptaddress, 
]{revtex4-1}

\usepackage[T1]{fontenc}

\usepackage[normalem]{ulem}
\usepackage{amsthm,nccmath}

\usepackage{xcolor}

\usepackage{hyperref}
\usepackage{float}

\usepackage{tabularx}

\hypersetup{
	colorlinks,
	linkcolor={blue!90!black}, 	
	citecolor={red!90!black},
	urlcolor={blue!85!black}
}

\usepackage{amsmath}
\usepackage{amssymb}
\usepackage{graphicx}

\usepackage{mathtools}

\usepackage{wasysym}

\begin{document}


\title{Dynamical control of the conductivity of an atomic Josephson junction}

\author{Beilei Zhu}
\email{bzhu@physnet.uni-hamburg.de}
\affiliation{Zentrum f\"ur Optische Quantentechnologien and Institut f\"ur Laserphysik, Universit\"at Hamburg, 22761 Hamburg, Germany}

\author{Vijay Pal Singh}
\affiliation{Zentrum f\"ur Optische Quantentechnologien and Institut f\"ur Laserphysik, Universit\"at Hamburg, 22761 Hamburg, Germany}

\author{Junichi Okamoto}
\affiliation{Institute of Physics, University of Freiburg, Hermann-Herder-Str. 3, 79104 Freiburg, Germany}
\affiliation{EUCOR Centre for Quantum Science and Quantum Computing, University of Freiburg, Hermann-Herder-Str. 3, 79104 Freiburg, Germany}

\author{Ludwig Mathey}
\affiliation{Zentrum f\"ur Optische Quantentechnologien and Institut f\"ur Laserphysik, Universit\"at Hamburg, 22761 Hamburg, Germany}
\affiliation{The Hamburg Centre for Ultrafast Imaging, Luruper Chaussee 149, 22761 Hamburg, Germany}

\date{\today}


\begin{abstract}

We propose to dynamically control the conductivity of a Josephson junction composed of two weakly coupled one dimensional condensates of ultracold atoms. A current is induced by a periodically modulated potential difference between the condensates, giving access to the conductivity of the junction. By using parametric driving of the tunneling energy, we demonstrate that the low-frequency conductivity of the junction can be enhanced or suppressed, depending on the choice of the driving frequency. The experimental realization of this proposal provides a quantum simulation of optically enhanced superconductivity in pump-probe experiments of high temperature superconductors. 

\end{abstract}

\maketitle


\section{Introduction}

Recently, light-induced or enhanced superconductivity has been discovered in superconducting materials such as YBa$_2$Cu$_3$O$_{6+x}$ \cite{Kaiser2014, HuW2014, Forst2014, Mankowsky2015, Mankowsky2017} or K$_3$C$_{60}$ \cite{Mitrano2016, Cantaluppi2018} using pump-probe techniques with mid-infrared lasers. This has triggered theoretical investigations of the origin and mechanism of optical control of superconductivity. Based on microscopic models, various mechanisms have been proposed such as enhancement of 
electron-phonon coupling \cite{Knap2016, Babadi2017, Kennes2017, Murakami2017,sentef2017}, control of competing order \cite{patel2016, sentef2016, ido2017, mazza2017, wang2018, bittner2019}, photo-induced $\eta$-pairing \cite{kaneko2020, kaneko2019a,li2019a} and cooling in multi-band systems \cite{nava2018, werner2018}. Meanwhile, phenomenological approaches have been used to understand the effect of superconducting fluctuations in photo-excited systems \cite{denny2015, hoppner2015, okamoto2016, okamoto2017, schlawin2017, chiriaco2018, harland2019, iwazaki2019, lemonik2019}. In Refs. \cite{hoppner2015, okamoto2016, okamoto2017}, we proposed a mechanism based on parametrically driven Josephson junctions for light-enhanced superconductivity. This mechanism is also reflected in a redistribution of current fluctuations, such that the low-frequency part of the system is effectively cooled down leading to an enhancement of the inter-layer tunneling energy, see Ref. \cite{hoppner2015}.

Given the complexities of light-induced dynamics in strongly correlated solids, it is conceptually instructive to explore proposed mechanisms in a well-defined physical system, which isolates specific features of the solid state system. In particular, cold atom systems are highly tunable model systems, that provide toy models for more complex systems, in the spirit of quantum simulation. In this paper, we will utilize the ability of cold atom technology to design and control Josephson junction systems.  Atomic Josephson junctions \cite{Cataliotti2001,  Albiez2005, Levy2007, LeBlanc2011,  Betz2011, Spagnolli2017, Valtolina1505} have been realized experimentally to study coherent transport \cite{Chien2015, Krinner2017}, atomic conductivity \cite{Anderson2019}, and the dynamics of Josephson junctions of two dimensional cold atomic gases \cite{Luick2020,Singh2020}. This provides an ideal platform to simulate the dynamics of a parametrically driven Josephson junction.
\begin{figure*}
\includegraphics[scale=0.62]{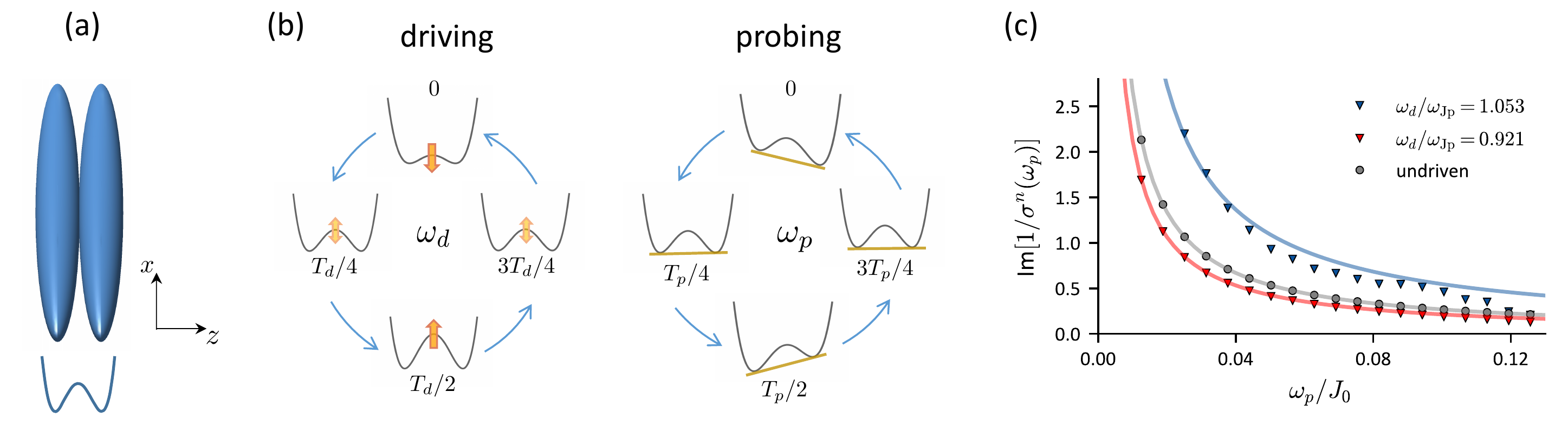}
\caption{\label{fig:fig1}
(a) A Josephson junction composed of two weakly coupled 1D condensates in a double-well potential. 
(b) Illustration of the driving and probing process, implemented via periodic modulation of the barrier height and the potential difference between the two potential minima, with driving (probing) frequency $\omega_d$ ($\omega_p$). 
(c) We depict the imaginary part of the inverse of the conductivity of the atomic Josephson junction, Im$[1/\sigma^n(\omega)]$, obtained from our numerical classical field simulations. It displays a $1/\omega$ divergence, reminiscent of the conductivity of a junction of charged particles. For parametric driving with a blue-detuned driving frequency of $ \omega_d / \omega_{\mathrm{Jp}} = 1.053 $, the prefactor of the divergence is enhanced by $ \sim 104\% $. For red-detuned driving at $\omega_d / \omega_{\mathrm{Jp}} = 0.921 $, the prefactor is reduced by $ \sim 21\% $. This demonstrates dynamical control of conductivity of an atomic junction, based on the mechanism put forth in Ref. \cite{okamoto2016}. }
\end{figure*}

In this paper, we propose to perform dynamical control of the conductivity of a Josephson junction composed of two weakly coupled one dimensional (1D) condensates, see Fig.~\ref{fig:fig1}(a). This proposal is motivated by the mechanism of parametrically enhanced conductivity, that we established in Refs.~\cite{okamoto2016, okamoto2017}. For that purpose, two dynamical processes have to be introduced in the system of coupled condensates: One is the analogue of the probing process, and the second one is the analogue of the pumping process or optical driving. 
As shown in Fig.~\ref{fig:fig1}(b), we implement driving and probing via periodical modulation of the tunneling energy and the potential difference between the condensates, respectively.
The probe induces a current, allowing us to determine the conductivity $\sigma^n(\omega)$ of the junction of neutral atoms. 
To serve as a quantum simulation for a Josephson junction of charged particles, we will determine the relation of $\sigma^n(\omega)$ to the conductivity $\sigma^c(\omega) $ of a junction of charged particles below, where we demonstrate that $\sigma^n(\omega)$ is inversely proportional to $\sigma^c(\omega)$. 
This interpretation derives from the difference of a U(1) symmetry of a system of neutral particles and a U(1) gauge symmetry of a system of charged particles. 
Using classical field simulations, we show that the density imbalance between the condensates is suppressed at low probe frequencies when the parametric driving frequency is above the Josephson plasma frequency, and enhanced below the plasma frequency, which constitutes dynamical control of conductivity. 
Based on a two-site Bose-Hubbard (BH) model, we derive analytical expressions for $1/\sigma^n(\omega)$ for an undriven and a driven system. The comparison between the analytical estimates and the simulations shows good agreement.
We note that while the key physics occurs in the motion of the relative phase between the condensates, the 1D geometry of the two subsystems of the junction serves as an entropy bath, which slows down the heating of the system. This reduced heating rate of the system allows for a long probing time used in this proposal. 
Finally, we relate the dynamical renormalization of the conductivity of the atomic junction to a resistively and capacitively shunted junction (RCSJ) model, utilized in the description of electronic circuits. This can be visualized as a dynamically driven washboard potential.
We note that similar models have been studied in Refs. \cite{Smerzi1997, Paraoanu2001, Meier2001, Gati2007, Spuntarelli2007, Boukobza2010, chen2020}. Here, we present how parametric driving of the junction near its resonance frequency affects its conductivity, which constitutes the key insight of our study. 

This paper is organized as follows.
In Sec.~\ref{sec:model}, we describe the numerical simulation method for the system and show  the numerical results for the time evolution of the density imbalance between the two condensates for an undriven system and a driven system, which demonstrate that parametric driving affects the response at low probing frequencies.
In Sec.~\ref{sec:conductivity}, we derive an analytical estimate of the conductivity $\sigma^n(\omega)$ of neutral particles using a two-site BH model.
Furthermore, we derive the conductivity $\sigma^c(\omega)$ of a junction of charged particles, and discuss the relation of $\sigma^n(\omega)$ and $\sigma^c(\omega)$.
In Sec.~\ref{sec:MC} we show the numerical results that demonstrate dynamical control of the conductivity of a junction composed of two coupled 1D condensates.
In Sec.~\ref{sec:toy model}, we expand on the analytical approach of Sec.~\ref{sec:conductivity}, and derive how parametric driving renormalizes the conductivity $\sigma^n(\omega)$ of a junction of neutral particles. Furthermore, we compare this analytical prediction with our numerical results.
In Sec.~\ref{sec: driving mechanism}, we relate the analytical result for a junction of neutral particles, derived in Sec.~\ref{sec:toy model}, to a parametrically driven junction of charged particles. We conclude in Sec.~\ref{sec:conclusion}.
%
%
%
%
\section{Simulation method} 
\label{sec:model}
We consider a Josephson junction composed of two 1D condensates, as shown in Fig.~\ref{fig:fig1}(a). We numerically simulate the dynamics of this system using the classical field method of Ref. \cite{Singh2016}. For the numerical implementation, we represent the system of two coupled condensates, as a lattice model, which takes the form of a BH Hamiltonian 
\begin{eqnarray}
\hat{H_0}  &= -\sum_{\langle \alpha, \beta \rangle} J_{\alpha \beta}  \hat{b}_{\alpha}^{\dagger} \hat{b}_{\beta}  + \frac{U}{2} \sum_{\alpha}  \hat{n}_{\alpha}( \hat{n}_{\alpha} - 1) .  
\label{eq:Hamil}
\end{eqnarray}
$\hat{b}_{\alpha}^{\dagger}$ $(\hat{b}_{\alpha})$ is the bosonic creation (annihilation) operator at site $\alpha$. $\langle \alpha, \beta \rangle$ denotes nearest neighbour sites $\alpha$ and $\beta$. The lattice dimensions are $N_x \times N_z$, where we choose $N_x=50$ and $N_z=2$. 
Each site index $\alpha=(i,j)$ encodes the two coordinates $i$ and $j$, with $i \in [1,50]$ and $j=1,2$. $\hat{n}_{\alpha} = \hat{b}_{\alpha}^{\dagger} \hat{b}_{\alpha} $ is the number operator at site $\alpha$. Along the \textit{z-} direction, the tunneling energy $J_{\alpha \beta}$ is given by $J_z$, which is the tunneling energy of the double-well potential. 
In the undriven system, this tunneling energy is a constant, $J_z = J_0$. 
We will use $J_0$ as the energy scale throughout this paper. 
To capture the continuous 1D condensates numerically, we discretize the motion along the \textit{x-}direction, with a discretization length $l_x$.
This results in an effective tunneling energy $J_x = \hbar^2 / (2 m l_x^2)$, where $m$ is the atomic mass and $\hbar$ is the reduced Planck constant \footnote{ $l_x$ is chosen to be smaller than the healing length  $\xi= \hbar/\sqrt{2 m g_{\mathrm{1D}} n}$ and the thermal de Broglie wavelength $\lambda = \sqrt{2\pi \hbar^2/(m k_B T)}$, where $k_B$ is Boltzmann constant and $T$ is the temperature.}. 
\begin{figure*}[!t]
\includegraphics[width=1.0\textwidth]{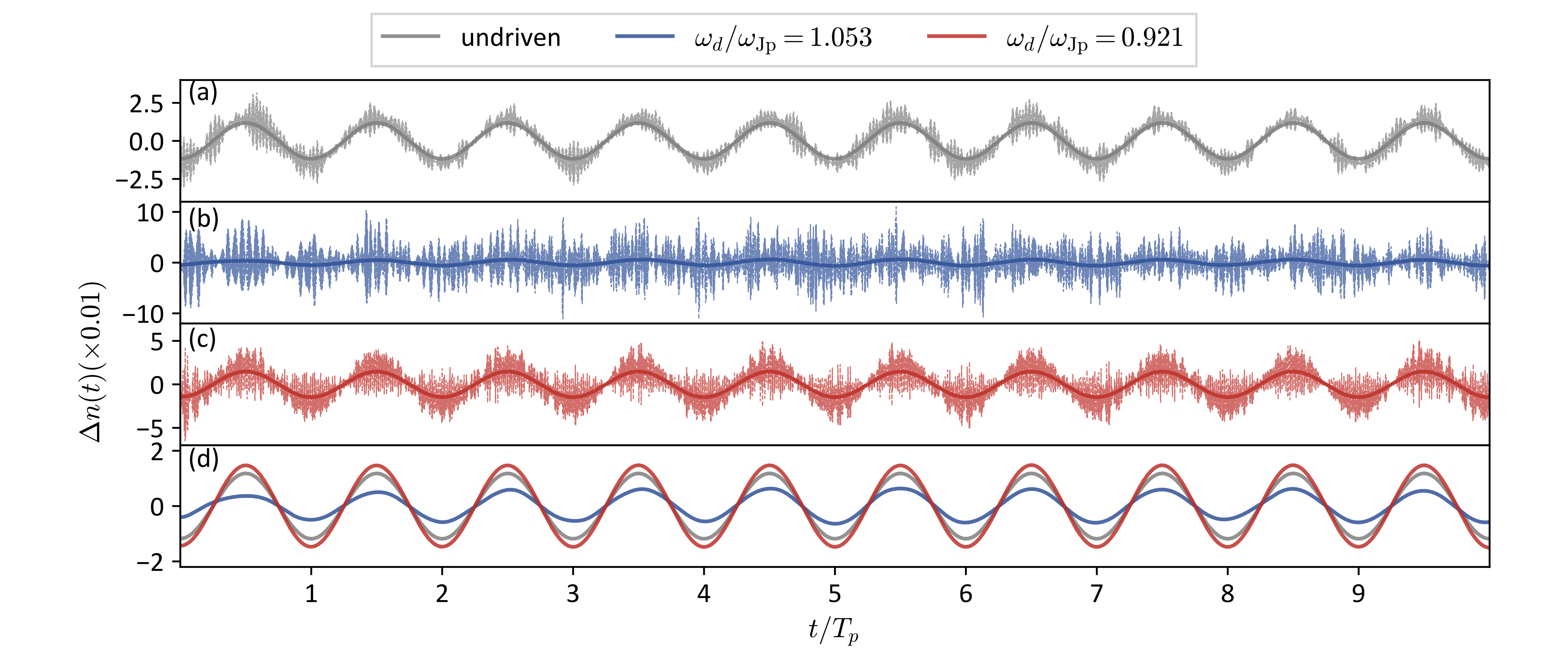}
\caption{\label{fig:time evolution} Density imbalance $\Delta n(t)$ as a function of time $t$. The system is subjected to the probing term of Eq.~(\ref{eq:probe Hamiltonian}) with an amplitude $V_0=0.01 J_0 $ and a frequency $\omega_p = 2\pi\times 0.002 J_0 $. Time $t$ is displayed in units of the probing period $T_p = 500 J_0^{-1}$. 
The dashed lines in (a)-(c) are $\Delta n(t)$, averaged over 500 realizations. The continuous lines are the low-frequency filtered $\Delta n(t)$, where we use a Gaussian filter with a time scale of $0.1 T_p$, see text. The low-frequency filtered $\Delta n(t)$ of (a)-(c) is displayed in (d) as well, for comparison. The probing frequency is significantly smaller than the resonance frequency of the junction, which is $\omega_{\mathrm{Jp}} = 2.28 J_0 $. In (a) we display $\Delta n(t)$ of the undriven system. In (b) and (c) we drive the system parametrically, see Eq.~(\ref{eq:driving term}). In (b) the driving frequency is $\omega_d / \omega_{\mathrm{Jp}} = 1.053 $, and therefore blue-detuned, in (c) it is $\omega_d / \omega_{\mathrm{Jp}} = 0.921 $, and therefore red-detuned. The comparison in (d) demonstrates dynamical control of the low-frequency response of the junction.}
\end{figure*}
In this discretized representation, the on-site repulsive interaction is determined by  $U=g_{\mathrm{1D}}/ l_x $, where $g_{\mathrm{1D}} = 2 \hbar^2 a_s/(m a_y a_z)$. $a_{s}$ is the \textit{s-}wave scattering length and $a_y$ $(a_z)$ is the oscillator length due to trap confinement along \textit{y-} (\textit{z-}) direction.
In the following, we set $\hbar = 1$.
We use $U = 0.15 J_0 $ throughout this paper. In our classical field representation, we replace the operators in Hamiltonian (\ref{eq:Hamil}) and its corresponding equations of motion by complex numbers. We sample initial states from a grand canonical ensemble with chemical potential $\mu$ and temperature $T$. We choose $J_x = 3.3 J_0$ and $T = 0.2 J_0/k_B$, $k_B$ being the Boltzman constant, and adjust $\mu$ such that the density per site is $n_0 = 2 $. 
For the probe, we add the following term
\begin{eqnarray}
	\hat{H}_{\text{pr}} =  V(t) \cdot \Delta \hat{N} ,
	\label{eq:probe Hamiltonian}
\end{eqnarray}
where the number imbalance $\Delta \hat N $ is 
\begin{eqnarray}
	\Delta \hat{N}  = \frac{1}{2} \sum_i \left( \hat{n}_{(i,1)} - \hat{n}_{(i,2)} \right)
\end{eqnarray}
with $V(t ) = V_{0}\cos (\omega_{p}t ).$ $V_0$ is the probe amplitude and $\omega_{p}$ the probe frequency. 
The oscillating potential induces an oscillating current and density motion between the condensates, which we use to determine the conductivity, as we describe below.
To simulate parametric driving, we modulate the tunneling energy $J_z$ as
\begin{eqnarray}
	J_z(t) = J_{0} \left[ 1+ A \cos \left( \omega_d t \right)  \right] \label{eq:driving term},
\end{eqnarray}
where $A$ is the driving amplitude and $\omega_{d}$ the driving frequency. As a key quantity to determine the conductivity $\sigma^n(\omega)$, we calculate the density imbalance, averaged over each 1D condensate, 
\begin{eqnarray}
	\Delta n (t) \equiv  \frac{ \langle \Delta \hat{N} (t) \rangle }{ N_x } =  \frac{2 n_0 }{ N } \langle \Delta \hat{N} (t) \rangle 
\end{eqnarray}
where $\langle ... \rangle $ denotes the average over the thermal ensemble and $N
$  is the total particle number in the system.

In Fig.~\ref{fig:time evolution} we present an example that demonstrates the main physical effect that we propose to realize experimentally. 
We show the time evolution of the density imbalance $\Delta n (t)$, averaged over 500 trajectories, as a function of time. 
The system of coupled condensates is subjected to a probing term with $V_0 = 0.01 J_0 $ and a small probing frequency of $\omega_p = 2 \pi / T_p =  2 \pi \times 0.002 J_0$. 
The probing period $T_p$ is used as a time scale in Fig.~\ref{fig:time evolution}. 
In Fig.~\ref{fig:time evolution}(a), we show  $\Delta n (t)$ for an undriven system, which displays high frequency fluctuations due to thermal noises. We filter these fluctuations using a filter function with Gaussian kernel $g(t) = \exp(- t^2 / \sigma_t^2 )$. 
We choose the time scale $\sigma_t =T_p/10$. The low-frequency part of the motion of the density imbalance displays a periodic motion at the probing frequency $\omega_p$. 
It is depicted in Fig.~\ref{fig:time evolution}(a), in addition to the unfiltered data, and also in Fig.~\ref{fig:time evolution}(d), to be compared to the density motion of a parametrically driven system, as we describe below.

We now calculate $\Delta n (t)$ for a driven junction.  We drive the tunneling energy between the condensates as described by Eq.\eqref{eq:driving term}. We use the driving amplitude $A=0.4$. 
In Figs.~\ref{fig:time evolution}(b) and (c), we show $\Delta n(t)$ for a blue-detuned ($ \omega_d/\omega_{\mathrm{Jp}} = 1.053 $) and a red-detuned ($\omega_d / \omega_{\mathrm{Jp}} = 0.921$) driving frequency, respectively. 
$\omega_{\mathrm{Jp}}$ is the Josephson plasma frequency, which we estimate as $\omega_{\mathrm{Jp}}  = \sqrt{4 J_0(J_0 + U n_0)} $, as we describe below. 
For the parameter choice of this example, we have $\omega_{\mathrm{Jp}}  = 2.28 J_0$, which we use as a frequency scale for the driving frequency. 
We note that this choice of the driving amplitude and driving frequency is outside of the parametric heating regime, which allows for a long driving time. 

As depicted in Figs.~\ref{fig:time evolution} (b) and (c), we also determine the low-frequency filtered density imbalance, which we calculate via Gaussian filtering as described above.
We compare these averaged values in Fig.~\ref{fig:time evolution}(d). The amplitude of the oscillation of the density imbalance is increased due to parametric driving with a red-detuned driving frequency, and decreased due to driving with a blue-detuned frequency. This observation exemplifies the main point of parametric control of conductivity, for an atomic Josephson junction. For red-detuned driving, the low-frequency limit of the response to a potential difference between the two condensates is increased. To achieve the same response statically, a larger tunneling energy would be required. This dynamically induced behaviour is therefore parametrically enhanced superfluidity.
For blue-detuned driving, the amplitude of the oscillation of the density imbalance is reduced, which indicates a reduction of superfluidity. 
This constitutes the essence of control of conductivity via parametric driving. We elaborate on this observation below and relate it to the conductivity of a parametrically driven junction of charged particles. As we demonstrate, the frequency dependence is inverted: For blue-detuned parametric driving, the superconducting response is enhanced, for red-detuned driving the response is reduced.

%

\section{conductivity}
\label{sec:conductivity}

In this section, we derive the conductivity of a Josephson junction of neutral particles and of charged particles, at linear order. The resulting expressions are proportional to the inverse of each other, as we discuss below.

To provide an estimate for the conductivity of an atomic junction, we consider a two-site BH model in phase-density representation, in linearized form:
 \begin{eqnarray}
H_2 = \left( \frac{J_0}{n_{0}} + U \right) \Delta n^{2} + J_0 n_{0} \theta^{2} + V(t) \Delta n . \label{eq:H2}
\end{eqnarray}
$\theta$ is the phase difference of the two condensates. $\Delta n = (n_1 - n_2)/2 $ is the density imbalance. The equations of motion are
\begin{eqnarray}
	\Delta \dot n &=& 2 J_0 n_{0}\theta, \label{eq:eom1}\\
	\dot \theta  &=& -2 \left( \frac{J_0}{n_{0}}+U \right) \Delta n - V (t ). \label{eq:eom2}
\end{eqnarray}
Eliminating $\Delta n$, we obtain an equation of motion for $\theta$,
\begin{eqnarray}
\ddot {\theta } + \gamma \dot{ \theta} + \omega_\mathrm{Jp}^2 \theta = -\dot{ V} (t)  ,
\label{eq:theta 2nd}
\end{eqnarray}
where $\gamma$ is included phenomenologically to describe damping. $\omega_{\mathrm{Jp}}  = \sqrt{4 J_0(J_0 + U n_0)} $ is the Josephson plasma frequency, as stated in Sec.~\ref{sec:model}. The Fourier transform of Eq.~(\ref{eq:theta 2nd}) can be written as
\begin{eqnarray}
\theta(\omega) =  \frac{  - i \omega V(\omega)} { \omega^2 - \omega_\mathrm{Jp}^2  + i \gamma \omega}, \label{eq:theta solution}
\end{eqnarray}
which relates the phase to the external probing potential. The particle current is determined by $j = - \Delta \dot n $.
The minus sign is explicitly included to specify that a positive $j$ refers to a current flowing from condensate $1$ to condensate $2$, and a negative $j$ to the opposite direction.
The conductivity of a junction of neutral particles is defined as $\sigma^n(\omega) \equiv j(\omega)/V(\omega)$.
Combining Eqs.~(\ref{eq:eom1}) and (\ref{eq:theta solution}), we obtain 
\begin{eqnarray}
\sigma^n(\omega) = 2 J_0 n_0 \frac{ i \omega } { \omega^2 - \omega_\mathrm{Jp}^2  + i \gamma \omega }.
\label{eq: sigma n} 
\end{eqnarray}
So the conductivity of an atomic junction is a Lorentian with its maximum at the resonance frequency $\omega_{\mathrm{Jp}}$, multiplied by the frequency $\omega$.

To develop the relation between the transport across an atomic junction and a junction of charged particles, we derive the conductivity of the RCSJ model of a junction. The linearized equation of motion for the phase difference of a Josephson junction is \cite{okamoto2016} 
\begin{eqnarray}
	\ddot \phi + \gamma_c \dot{\phi} + \omega_{\text{Jp,c}}^{2} \phi = \tilde{I},
	\label{eq:sG eq}
\end{eqnarray}
where $ \tilde{I} \equiv \omega_{\text{Jp,c}}^2 I/ J_0^c $, with $ I $ being the external current. $\omega_{\text{Jp,c}} = (2e/\hbar)^2 J_0^c/C $ is the Josephson plasma frequency, where  $J_0^c$ is the bare Josephson coupling energy and $C$ is the capacity determined by the geometry of the junction. The conductivity is defined as $ \sigma^c(\omega) \equiv  I d / V_c $, where $d$ is the distance between the superconductors. The voltage difference across the junction is given by the Josephson relation, $V_c = \frac{\hbar}{2e}\dot{\phi}$, where $e$ is the charge of an electron.
We then obtain  
\begin{equation}
\sigma^c ( \omega )  = \frac{\hbar}{2e}\frac{C  d}{i\omega } \left( \omega^{2} - \omega_\text{Jp,c}^{2} + i \gamma_c \omega \right). \label{eq: sigma c}
\end{equation}

Comparing Eqs.~(\ref{eq: sigma n}) and (\ref{eq: sigma c}), we observe that the conductivity $\sigma^n(\omega)$ and $\sigma^c(\omega)$ are proportional to each other's inverse, i.e.,  
\begin{eqnarray}
1/\sigma^n(\omega) \sim  \sigma^c(\omega).
\label{eq:sigma theta}
\end{eqnarray}
This relation motivates us to display $1 / \sigma^n(\omega)$ throughout this paper, for example in Fig.~\ref{fig:fig1}(c) and Fig.~\ref{fig:delta-n}.  This quantity features a $ 1 / \omega$ divergence in its imaginary part, and a zero crossing at the resonance frequency, and therefore directly resembles the conductivity of a junction of charged particles.

The origin of this relation derives from a comparison of Eq.~(\ref{eq:theta 2nd}) and (\ref{eq:sG eq}).
In both cases, the equations have the form of a driven oscillator. 
This results in a linear relation between the current and the potential, when expressed in frequency space. 
The phase of the atomic junction relates to the current, at linear order, and is therefore the quantity that responds to the external perturbation $ - \dot V(t) $ in Eq.~(\ref{eq:theta 2nd}). 
However, for the electronic junction, the phase is related to the external potential, due to the gauge theoretical relation of phase and vector field, whereas the inhomogeneous term in Eq.~(\ref{eq:sG eq}) is the current. Therefore the roles of current and external potential are reversed between Eq.~(\ref{eq:theta 2nd}) and Eq.~(\ref{eq:sG eq}), resulting in the inverse response function.

\section{Numerical results}
\label{sec:MC}

\begin{figure}
	\includegraphics[scale=1.1]{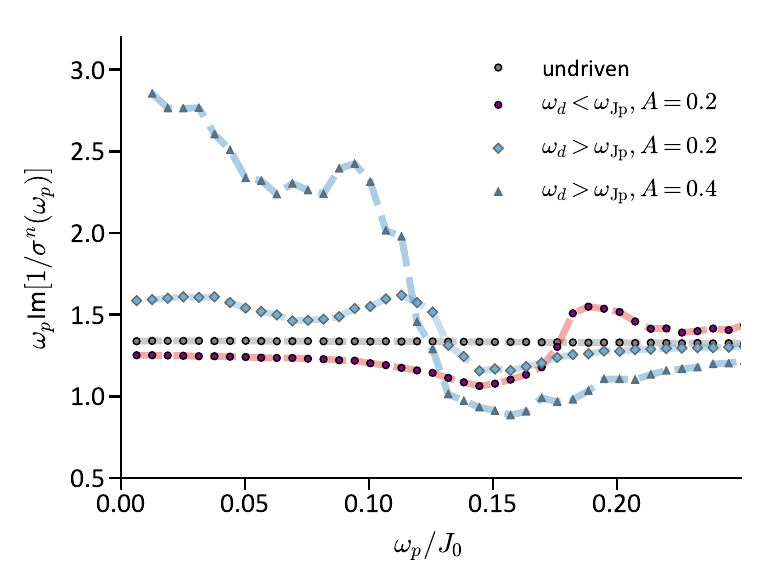}
\caption{ \label{fig:sigma} Numerical simulation results for $ \omega_p \text{Im}[1/ \sigma^n(\omega_p)]$  of the undriven system (grey circles), for red-detuned driving (purple circles) with $\omega_d / \omega_{\mathrm{Jp}} = 0.921 $, and for blue-detuned driving with $\omega_d / \omega_{\mathrm{Jp}} = 1.053 $ and driving amplitudes $ A = 0.2 $ (diamonds) and $ A = 0.4$ (triangles). }
\end{figure}
We present how the inverse of the conductivity $1/\sigma^n(\omega)$ is affected by parametric driving at a blue-detuned driving frequency of $\omega_d / \omega_{\mathrm{Jp}} = 1.053 $ and a red-detuned driving frequency of $\omega_d / \omega_{\mathrm{Jp}} = 0.921$.
In Fig.~\ref{fig:sigma}, we show $\omega_p \text{Im} [1 / \sigma^n ( \omega_p ) ]$ in the low probing frequency regime. For each $\omega_p$, we determine the time evolution of $ \Delta n(t) $ over a time duration of $10 T_p $, and extract the Fourier component $\Delta n(\omega_p)$.  
For the undriven system, $\omega_p \text{Im}[1/\sigma^n(\omega_p)]$ approaches a constant value of $\sim \omega_{\mathrm{Jp}}^2/(2 J_0 n_0)$ in the limit of $\omega_p \to 0$.
In the presence of parametric driving, the low frequency response is modified. When the driving frequency is larger than the Josephson plasma frequency, i.e., $\omega_{d} > \omega_{\mathrm{Jp}}$, the quantity $\omega_p \text{Im}[1/\sigma^n(\omega_p)]$  is enhanced for $\omega_p < \omega_d - \omega_{\mathrm{Jp}}$, indicating a reduced effective tunneling energy across the junction. 
The magnitude of enhancement depends on the driving amplitude, as shown in Fig.~\ref{fig:sigma}. For larger driving amplitude $A$, $\omega_p \text{Im}[1/\sigma^n(\omega_p)]$ shows a larger enhancement. Above the frequency difference, i.e., $\omega_p > \omega_d - \omega_{\mathrm{Jp}}$, the quantity $\omega_p \text{Im}[1/\sigma^n(\omega_p)]$ is reduced. 
This observation that the enhancement of $\omega_p \text{Im}[1/\sigma^n(\omega_p)]$ at low probing frequency limit is accompanied by the reduction of $\omega_p \text{Im}[1/\sigma^n(\omega_p)]$ above the frequency difference, i.e., $\omega_p > \omega_d - \omega_{\mathrm{Jp}}$, is reminiscent of the redistribution of phase fluctuations described in \cite{hoppner2015}.  On the other hand, for a red-detuned driving frequency, i.e., $\omega_d < \omega_{\mathrm{Jp}}$ , $\omega_p \text{Im}[1/\sigma^n(\omega_p)]$ is reduced for $\omega_p < \omega_d - \omega_{\mathrm{Jp}} $ while increased for $\omega_p > \omega_d - \omega_{\mathrm{Jp}} $.  We note that the enhancement and suppression effect is most pronounced for $\omega_d$ close to $\omega_{\mathrm{Jp}}$. For $\omega_d$ far away from $\omega_{\mathrm{Jp}}$, the effect of enhancement and suppression is diminished.

\section{ Parametric control of conductivity }
\label{sec:toy model}

\begin{figure}
	\includegraphics[scale=1.04]{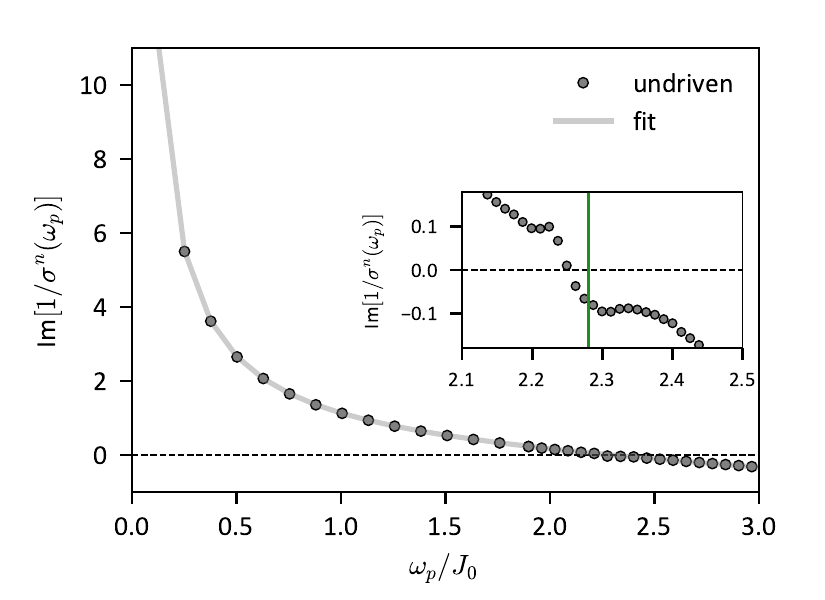}
\caption{\label{fig:delta-n} $\text{Im}[1/\sigma^n(\omega_p)]$ as a function of the probing frequency $\omega_p$ for an undriven system, obtained from the numerical simulation (circles) and analytical prediction of Eq.~(\ref{eq: sigma th}). The probing amplitude is $ V_0 = 0.01 J_0 $. We obtain the fitting parameters for the density $ n_0 = 1.8779 $, for the damping $ \gamma=0.05 J_0 $ and for the Josephson plasma frequency $ \omega_{\mathrm{Jp}} = 2.2939 J_0 $. Inset: zoom-in near $\omega_{\mathrm{Jp}}$. The green vertical line indicates the analytical estimate of the Josephson plasma frequency of $\omega_\mathrm{Jp} = 2.28 J_0$. }%
\label{fig:undriven conductivity}
\end{figure}
%
Based on the analytical estimate of the conductivity that we presented in Sec.~\ref{sec:conductivity}, the inverse of the conductivity is 
\begin{equation}
1/\sigma^n ( \omega )  = \frac{1}{2 J_0 n_0}\frac{1}{i\omega } \left( \omega^{2} - \omega_\mathrm{Jp}^{2} + i \gamma \omega \right). \label{eq: sigma th}
\end{equation}
In Fig.~\ref{fig:undriven conductivity}, we show the numerical results for $\text{Im}[1/\sigma^n(\omega_p)]$. It displays a $1/\omega$ divergence in the low frequency regime, that is associated with the low-frequency behaviour of the conductivity of a superconductor.
We fit the numerical data with formula \eqref{eq: sigma th}, which gives for the condensate density $n_0 = 1.8779$, for the damping $\gamma = 0.05 J_0$ and for the plasma frequency $\omega_{\mathrm{Jp}} = 2.2939 J_0$. 
We note that the value of $n_0$ is close to the value of the numerical simulations, and the value of $\omega_{\mathrm{Jp}}$ is close to the analytical estimate $\omega_\mathrm{Jp} \equiv \sqrt{4 J_0(J_0 + U n_0)} \approx 2.28J_0$.
The zero crossing of $\text{Im} [ 1/ \sigma^n ( \omega_p ) ] $ signifies the Josephson plasma frequency. 
Again, we find that the analytical estimate is close to the numerical finding. To indicate the magnitude of the deviation from the linearized estimate, we display  $\text{Im}[ 1/\sigma^n(\omega_p) ]$ in the vicinity of the resonance in the inset. Small deviations are visible around the resonance, where nonlinear contributions are noticeable, due to the large amplitudes of the motion near resonance.

We now determine how the conductivity $\sigma^n (\omega)$ is modified by parametric driving. This analysis is closely related to the analysis presented in Ref.~\cite{okamoto2016}. We replace $J_0$ by $J(t)$ in Eq.~(\ref{eq:H2}). The equation of motion for $\Delta n(t)$ is
\begin{eqnarray}
	\Delta \ddot n  = 2 \dot J (t ) n_{0} \theta + 2J(t) n_0 \dot{\theta} - \gamma \Delta \dot n ,
	\label{eq:ODE for deltan}
\end{eqnarray}
where we include damping term phenomenologically with a damping parameter $ \gamma $. We note that time dependence of $ J(t) $ contributes an additional term on the right-hand side, compared to Ref.~\cite{okamoto2016}. This term is of the form of a damping term as well, in which the damping rate is modulated in time.
The oscillatory time dependence of $J(t)$ couples the mode $\Delta n(\omega_p)$ at the probing frequency to the modes $\Delta n(\omega_p \pm m  \omega_d)$, where $m \in \mathbb{Z}$. 
Using a three mode expansion, we write $\Delta n (t) = \sum_{j} \Delta n(\omega_j ) \exp(i \omega_j t)$, where $\omega_j = \omega_p + j \omega_d$ with $ j = 0 , \pm 1 $. 
The full expression for $1/\sigma^n(\omega)$ is given in Eq.~(\ref{eq:three mode}). In the limit of $\omega_p \to 0$, we obtain  
\begin{widetext}
\begin{eqnarray}
	\lim_{\omega_p \rightarrow 0}  \mathrm{Im}[ \omega_p / \sigma^n ( \omega_p )] & =  \frac{1}{2 J_0 n_0 } \frac{ A^2 \left (4 J_0^2 + \omega_{\mathrm{Jp}}^2 \right )  ( \omega_d^2 - \omega_{\mathrm{Jp}}^2 ) \left( 4 J_0^2 - \omega_d^2 + \omega_{\mathrm{Jp}}^2 \right) - 2 \omega_{\mathrm{Jp}}^2 \left[ \gamma^2 \omega_d^2+\left( \omega_d^2 - \omega_{\mathrm{Jp}}^2 \right )^2 \right ]}{ A^2 ( \omega_d^2 - \omega_{\mathrm{Jp}}^2 )  \left(4 J_0^2 - \omega_d^2 + \omega_{\mathrm{Jp}}^2 \right ) - 2 \gamma^2  \omega_d^2 - 2 \left( \omega_d^2 - \omega_{\mathrm{Jp}}^2 \right)^2  }. \label{eq:sigma 3 mode}
\end{eqnarray}
\end{widetext}
This modified expression depends on the driving amplitude and the driving frequency, and the damping parameter $\gamma$. We note that $A$ appears in the denominator as well. This is due to the $\dot{\theta}$ term that couples the probe to $J(t)$, which in turn leads to a probe input of three modes at frequencies $\omega_p, \omega_p + \omega_d, \omega_p - \omega_d$. Using an expansion with more and more modes, we expect that the contribution of $A$ to the denominator to play a lesser role.

\begin{figure}
	\centering
	\includegraphics[scale=1.08]{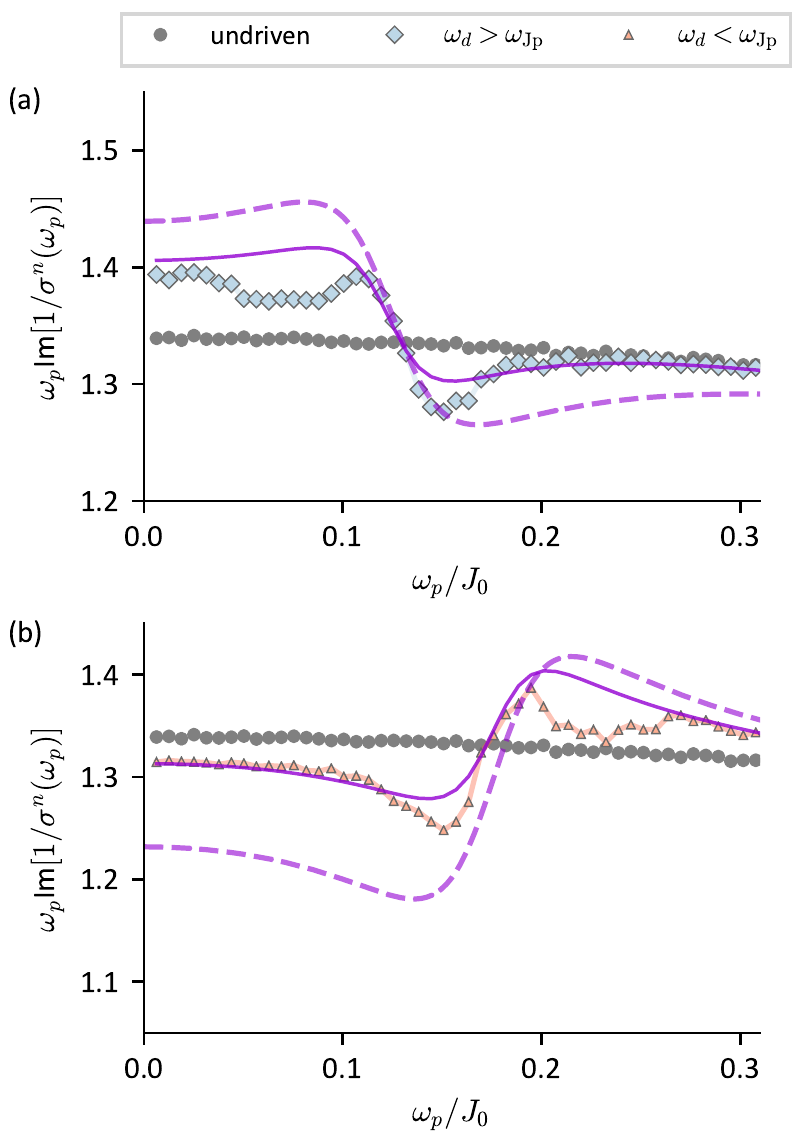}
\caption{\label{fig-sigma-dr}  Comparison of the numerical simulations (circles, diamonds, triangles) with the three mode expansion (purple dashed lines) of Eq.~(\ref{eq:three mode}) and the numerical solution (purple continuous lines) of Eqs.~(\ref{eq:eom1}) and (\ref{eq:eom2}) with the parameters $n_0 = 1.84 $ and $ \gamma = 0.06 J_0 $. The probe amplitude is $V_0 = 0.01 J_0 $. Driving amplitude is $ A = 0.1 $ and the driving frequency is $\omega_{d} / \omega_{\mathrm{Jp}} = 1.053$  for blue-detuned driving, and $\omega_{d} / \omega_{\mathrm{Jp}} = 0.921$ for red-detuned driving. }
\label{numerical comparison}
\end{figure}

In Fig.(\ref{fig-sigma-dr}), we compare the analytical result based on the three-mode expansion, Eq.~\eqref{eq:three mode}, the numerical result based on the Eqs.~(\ref{eq:eom1}) and (\ref{eq:eom2}), and the numerical simulation results of the two coupled condensates.
We show the case of blue-detuned driving, $\omega_d/\omega_{\mathrm{Jp}} = 1.053$ and the case of red-detuned driving, $\omega_d/ \omega_{\mathrm{Jp}} = 0.921$, both with $ A=0.1 $. We use the parameters $ n_0 = 1.84$ and $ \gamma = 0.06 J_0 $ for the three mode expansion and numerical result based on Eqs.~(\ref{eq:eom1}) and  (\ref{eq:eom2}). The numerical result of Eqs.~(\ref{eq:eom1}) and (\ref{eq:eom2}) matches the numerical simulation result well. The three mode expansion gives a qualitative estimate of enhancement and reduction of $1/\sigma^n(\omega)$. The overall shape of the response is that of a resonance pole located near $\omega_d - \omega_{\mathrm{Jp}} \approx 0.12 J_0$, broadened by the damping parameter $\gamma$, which depends on the temperature and nonlinear terms.

\section{Mechanism}
\label{sec: driving mechanism}
To describe the physical origin of the dynamical control effect that we present in this paper, we consider the equation of motion 
\begin{eqnarray}
	\ddot \phi + \gamma \dot{\phi} + \omega_{\mathrm{Jp}}^2 \left[ 1 +  A \cos \left( \omega_d t \right) \right] \sin \phi = I(t).
	\label{eq:driven sG eq}
\end{eqnarray}
This is the RCSJ model of a Josephson junction of charged particles, see Eq.~(\ref{eq:sG eq}), with an additional parametric modulation of the Josephson energy, see Ref.~\cite{okamoto2016}. Due to the similarity to an atomic Josephson junction, see e.g. Eqs.~(\ref{eq:theta 2nd}) and (\ref{eq:ODE for deltan}), this discussion provides an intuition for atomic junctions as well, with the re-interpretation of terms, discussed in Sec.~\ref{sec:conductivity}. 

\begin{figure}
\includegraphics[scale=0.72]{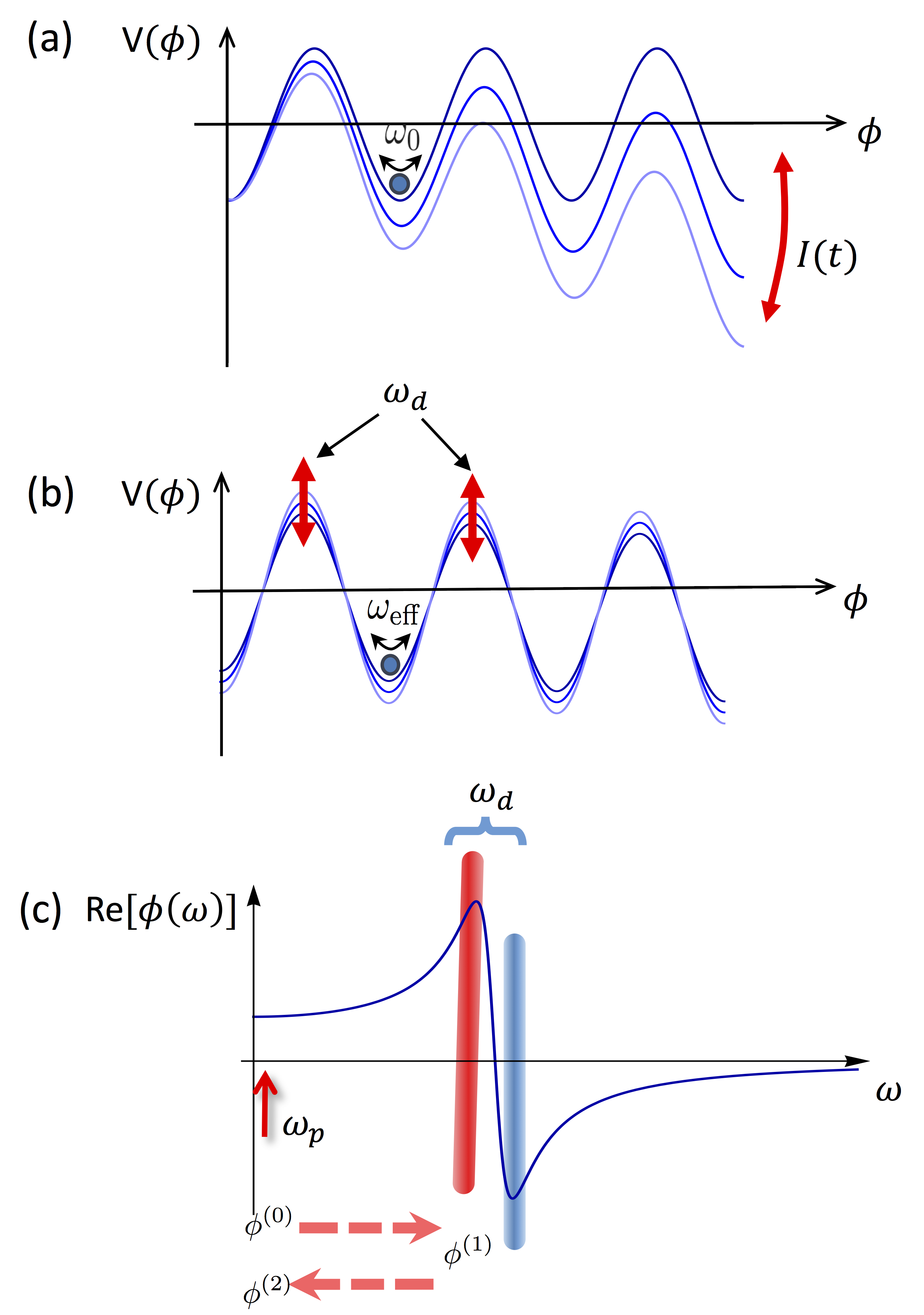}
\caption{ \label{washboard} In (a)  and (b) we depict the washboard potential representation of junction dynamics. In (a) we show the influence of an external current, in (b) we show the influence of parametric modulation of the junction energy. In (c), we show the response of the phase of the probe, and indicate the first and second order contributions to the modified response at $\omega_p$. }
\label{fig:washboard potential}
\end{figure}

Interpreted as a mechanical model, this equation describes a particle moving in a cosine potential, as depicted in Fig.~\ref{washboard}. This is the tilted-washboard potential representation of the RCSJ model \cite{Tinkham2004} with $ V(\phi) = - I(t) \phi - \omega_{\mathrm{Jp}}^2 [ 1 + A \cos(\omega_d t )] \cos \phi $.
The probe current $I(t)$ plays the role of tilting the potential up- and downward. If the external potential oscillates in time, the washboard potential is modulated with a linear gradient, oscillating in time. The Josephson plasma frequency $\omega_{\mathrm{Jp}}$ is the frequency of a particle oscillating around a minimum. The parametric driving term, given by $A\cos(\omega_d t)$ corresponds to a modulation of the height of the potential in time, as shown in Fig.~\ref{washboard}(b).

To describe the origin of the renormalization of the low-frequency response, and its sign change for driving frequencies above and below the resonance frequency, we present a perturbative argument. This approach provides an estimate of the conductivity that corresponds to the result of the three-mode expansion, expanded to second order. We expand $\phi = \phi^{(0)} + A \phi^{(1)} + A^2 \phi^{(2)} + ...$ by treating the driving amplitude $A$ as the expansion parameter. Assuming small amplitudes of the phase oscillation, we linearize $\sin \phi$. Inserting the expansion series into Eq. (\ref{eq:driven sG eq}), we the obtain the  equations
\begin{eqnarray}
 	\ddot{\phi}^{(0)} + \gamma \dot{\phi}^{(0)} + \omega_{\mathrm{Jp}}^2 \phi^{(0)} &=& I(t), \label{eq:zeroth pert} \\
 	\ddot{\phi}^{(n)} +\gamma \dot{\phi}^{(n)} + \omega_{\mathrm{Jp}}^2 \phi^{(n)}  &=& -  \omega_{\mathrm{Jp}}^2 A \cos(\omega_d t) \phi^{(n-1)}  \label{eq:1st pert} 
 \end{eqnarray}
in zeroth and \textit{n}-th order in $A$, respectively. We note that the $n$-th order solution is multiplied with $ \cos ( \omega_d t ) $ to provide the source term for the $(n+1)$-th order. With a monochromatic probe current $I(t) = I_0 \text{e}^{-i \omega_p t}$, the solutions to Eqs. (\ref{eq:zeroth pert}) and (\ref{eq:1st pert}), up to second order contributions, are
\begin{eqnarray}
	\phi^{(0)} &=&  \frac{ I_0  \text{e}^{-i \omega_p t} }{ \omega_{\mathrm{Jp}}^2 - \omega_p^2 - i \gamma \omega_p }, \label{phi 0} \\
	\phi^{(1)} &=& \frac{ - \omega_{\mathrm{Jp}}^2  (A / 2)  \phi^{(0)}   \text{e}^{-i \omega_d t} }{\omega_{\mathrm{Jp}}^2 - (\omega_p + \omega_d)^2 - i \gamma ( \omega_p + \omega_d ) }, \label{phi 1} \\
	\phi^{(2)} &=& \frac{ - \omega_{\mathrm{Jp}}^2 (A / 2) \phi^{(1)} \text{e}^{i \omega_d t}  }{\omega_{\mathrm{Jp}}^2 - \omega_p ^2 - i \gamma \omega_p}. \label{phi 2}
\end{eqnarray}
Each solution is the solution of a driven harmonic oscillator, responding to an external driving term. $\phi^{(0)}$ oscillates at frequency $\omega_p$ determined by the probe current $I(t)$, as indicated in Eq.~(\ref{eq:zeroth pert}).  
In the solution for $\phi^{(1)}$, $ - \omega_{\mathrm{Jp}}^2 A \cos(\omega_d t) \phi^{(0)} $ is the source term, and determines that $\phi^{(1)}$ oscillates at frequency $\omega_d + \omega_p$. 
If $\omega_d$ is close to the resonance frequency $ \omega_{\mathrm{Jp}} $, the amplitude of the response is large. 
As indicated in Fig.~\ref{washboard}(c), the motion of $\phi^{(1)}$ will pick up an additional phase of $\pi$, when the driving frequency is above the resonance frequency.
The second order correction $\phi^{(2)}$ oscillates at low frequency due to the oscillatory factor $\text{e}^{i \omega_d t} $. Therefore $\phi^{(2)} $ is the lowest order contribution to the motion at the probing frequency. 
The sign change at the resonance translates into $\phi^{(2)}$ having a positive or negative sign. 
Inserting Eqs.~(\ref{phi 0}) and (\ref{phi 1}) into Eq.~(\ref{phi 2}), in the limit of $\omega_p \to 0$, we obtain 
\begin{eqnarray}
\text{Re}[\phi^{(2)}] = \frac{ A^2 }{ 4 } \frac{\omega_{\mathrm{Jp}}^2 - \omega_d^2}{(\omega_d^2 - \omega_{\mathrm{Jp}}^2)^2 + \gamma^2 \omega_d^2 }.
\end{eqnarray}
Therefore, when the system is subjected to a blue-detuned driving, i.e., $\omega_d > \omega_{\mathrm{Jp}}$, the combined terms $\phi^{(0)}(\omega_p) + \phi^{(2)}(\omega_p)$ have a reduced magnitude, resulting in a stabilization of the phase. 
Similarly, for red-detuned driving, $\phi^{(2)}(\omega_p)$ has the same sign as $\phi^{(0)}(\omega_p)$, therefore the response of the phase is increased. 
For $\sigma^c(\omega_p)$ this implies that the conductivity is enhanced for blue-detuned driving and reduced for red-detuned driving, because the phase is proportional to electric field, while the current is held fixed. 
A reduction of the motion of the phase implies that the same current is induced with a smaller electric field, indicating an enhanced conductivity. 
For the conductivity of an atomic junction, the phase is proportional to the current, at linear order. So a reduction of the phase motion implies a reduction of the conductivity, which occurs at blue-detuned driving, while an increased current occurs at red-detuned driving, resulting in parametrically enhanced conductivity.

%

\section{conclusions}
\label{sec:conclusion}
We have demonstrated parametric enhancement and suppression of the conductivity of an atomic Josephson junction, composed of two weakly coupled 1D condensates. This is motivated by our proposed mechanism of parametric enhancement of the conductivity of light-driven superconductors \cite{okamoto2016}, which, in its simplest form, manifests itself in a single, parametrically driven Josephson junction. To demonstrate the analogous mechanism in a cold atom system, we discuss the relation between the conductivity of a junction of neutral particles and a junction of charged particles. We demonstrate that these are proportional to the inverse of each other. Based on this analogue, we propose to control the inverse of the conductivity of an atomic junction. We implement parametric control of the junction by periodic driving of the magnitude of the tunneling energy. We show numerically and analytically that the low-frequency limit of the inverse conductivity is enhanced for  parametric driving with a frequency that is blue-detuned with regard to the resonance frequency of the junction.  Similarly, the inverse of the conductivity is suppressed for parametric driving with a red-detuned frequency. This effect constitutes the central point of parametric enhancement of conductivity, which we propose to implement and verify in an ultracold atom system, which serves as a well-defined toy model, in the spirit of quantum simulation.
%

\section{acknowledgement}
\label{acknoledgement}

 This work was supported the DFG in the framework of SFB 925 and the excellence clusters ‘The Hamburg Centre for Ultrafast Imaging- EXC 1074 - project ID 194651731 and ‘Advanced Imaging of Matter - EXC 2056 - project ID 390715994.  B.Z. acknowledges support from China Scholarship Council (201206140012) and Equal Opportunity scholarship from University of Hamburg. J.O. acknowledges support from Research Foundation for Opto-Science and Technology and from Georg H. Endress Foundation.

\appendix%

\section{three mode expansion solution}
\label{sec:three mode}

To solve Eq.~(\ref{eq:ODE for deltan}), we first substitute $\theta$ and $\dot{\theta}$ using Eqs.~(\ref{eq:eom1}) and (\ref{eq:eom2}). A three mode expansion allows us to write 
\begin{eqnarray}
\Delta n (t)  &=& \Delta n(\omega_p) e^{ -i \omega_p t}  +\Delta n(\omega_p + \omega_d ) e^{ -i \omega_p t - i \omega_d t } \nonumber \\
		  &+& \Delta n( \omega_p - \omega_d ) e^{ -i \omega_p t + i \omega_d t} .
\end{eqnarray}
Now Eq.~(\ref{eq:ODE for deltan}) can be written in matrix form as
\begin{widetext}
\begin{gather}
  \begin{bmatrix}
   \omega_{Jp}^2 - i \gamma \Delta_{-d}  - \Delta{_d}^2  
   &   2 A J_0^2 + \omega_{\mathrm{Jp}}^2 A/2  
   & 0   \\
         2 A J_0^2 + ( \omega_{\mathrm{Jp}}^2 - \omega_d^2 ) A/2   
    &   \omega_{\mathrm{Jp}}^2 - i \gamma \omega_p  - \omega_p ^2  
    &    2 A J_0^2 + ( \omega_{\mathrm{Jp}}^2  - \omega_d^2 ) A/2  \\
   0 
   &  2 A J_0^2 + \omega_{\mathrm{Jp}}^2 A/2   
   &  \omega_{\mathrm{Jp}}^2 - i \gamma \Delta_{+d} - \Delta_{+d}^2   \\
   \end{bmatrix}
    \begin{bmatrix}
    \Delta n( \omega_p - \omega_d ) \\
    \Delta n( \omega_p ) \\
    \Delta n( \omega_p + \omega_d ) 
    \end{bmatrix}
    =
    J_0 n_0 V_0 \begin{bmatrix}
   -  A/2\\
    1 \\
   -  A/2
    \end{bmatrix}
\end{gather}
\end{widetext}
where we keep terms in $A$ up to first order. $\Delta_{+d} = \omega_p + \omega_d$ is the sum of the probing frequency $\omega_p$ and the driving frequency $\omega_d$.  $\Delta_{-d} = \omega_p - \omega_d$ is the difference frequency. With the solution of $\Delta n(\omega_p)$, we obtain the expression for the conductivity 
\begin{widetext}
\begin{eqnarray}
 1/\sigma^n(\omega_p) = \frac{ 2 \left\lbrace A^2 (4 J_0^2 + \omega_{\mathrm{Jp}}^2 ) B_{\text{fac}} /2 + \left[ \omega_{\mathrm{Jp}}^2 + i \omega_p (\gamma +i \omega_p ) \right] C_{\text{fac}} D_{ \text{fac} } \right\rbrace } 
 { J_0 n_0 \left(A^2  B_{\text{fac}} - 2  C_{\text{fac}}  D_{ \text{fac} } \right)}
 \label{eq:three mode}
\end{eqnarray} 
\end{widetext}
where $B_{\text{fac}} = \left(-i \gamma  \omega_p + \omega_d^2 - \omega_{\mathrm{Jp}}^2 + \omega_p^2 \right) \left(4 J_0^2 + \omega_{\mathrm{Jp}}^2 - \omega_d^2 \right) $, 
$ C_\text{fac} =  \omega_{\mathrm{Jp}}^2 + i \gamma \Delta_{-d} - \Delta_{-d}^2 $ 
and $ D_\text{fac} =  \omega_{\mathrm{Jp}}^2 + i \gamma \Delta_{+d} - \Delta_{+d}^2 $.


\bibliographystyle{apsrev4-1}

\bibliography{references.bib}

\begin{thebibliography}{54}%
\makeatletter
\providecommand \@ifxundefined [1]{%
 \@ifx{#1\undefined}
}%
\providecommand \@ifnum [1]{%
 \ifnum #1\expandafter \@firstoftwo
 \else \expandafter \@secondoftwo
 \fi
}%
\providecommand \@ifx [1]{%
 \ifx #1\expandafter \@firstoftwo
 \else \expandafter \@secondoftwo
 \fi
}%
\providecommand \natexlab [1]{#1}%
\providecommand \enquote  [1]{``#1''}%
\providecommand \bibnamefont  [1]{#1}%
\providecommand \bibfnamefont [1]{#1}%
\providecommand \citenamefont [1]{#1}%
\providecommand \href@noop [0]{\@secondoftwo}%
\providecommand \href [0]{\begingroup \@sanitize@url \@href}%
\providecommand \@href[1]{\@@startlink{#1}\@@href}%
\providecommand \@@href[1]{\endgroup#1\@@endlink}%
\providecommand \@sanitize@url [0]{\catcode `\\12\catcode `\$12\catcode
  `\&12\catcode `\#12\catcode `\^12\catcode `\_12\catcode `\%12\relax}%
\providecommand \@@startlink[1]{}%
\providecommand \@@endlink[0]{}%
\providecommand \url  [0]{\begingroup\@sanitize@url \@url }%
\providecommand \@url [1]{\endgroup\@href {#1}{\urlprefix }}%
\providecommand \urlprefix  [0]{URL }%
\providecommand \Eprint [0]{\href }%
\providecommand \doibase [0]{https://doi.org/}%
\providecommand \selectlanguage [0]{\@gobble}%
\providecommand \bibinfo  [0]{\@secondoftwo}%
\providecommand \bibfield  [0]{\@secondoftwo}%
\providecommand \translation [1]{[#1]}%
\providecommand \BibitemOpen [0]{}%
\providecommand \bibitemStop [0]{}%
\providecommand \bibitemNoStop [0]{.\EOS\space}%
\providecommand \EOS [0]{\spacefactor3000\relax}%
\providecommand \BibitemShut  [1]{\csname bibitem#1\endcsname}%
\let\auto@bib@innerbib\@empty
\bibitem [{\citenamefont {Kaiser}\ \emph {et~al.}(2014)\citenamefont {Kaiser},
  \citenamefont {Hunt}, \citenamefont {Nicoletti}, \citenamefont {Hu},
  \citenamefont {Gierz}, \citenamefont {Liu}, \citenamefont {Le~Tacon},
  \citenamefont {Loew}, \citenamefont {Haug}, \citenamefont {Keimer},\ and\
  \citenamefont {Cavalleri}}]{Kaiser2014}%
  \BibitemOpen
  \bibfield  {author} {\bibinfo {author} {\bibfnamefont {S.}~\bibnamefont
  {Kaiser}}, \bibinfo {author} {\bibfnamefont {C.~R.}\ \bibnamefont {Hunt}},
  \bibinfo {author} {\bibfnamefont {D.}~\bibnamefont {Nicoletti}}, \bibinfo
  {author} {\bibfnamefont {W.}~\bibnamefont {Hu}}, \bibinfo {author}
  {\bibfnamefont {I.}~\bibnamefont {Gierz}}, \bibinfo {author} {\bibfnamefont
  {H.~Y.}\ \bibnamefont {Liu}}, \bibinfo {author} {\bibfnamefont
  {M.}~\bibnamefont {Le~Tacon}}, \bibinfo {author} {\bibfnamefont
  {T.}~\bibnamefont {Loew}}, \bibinfo {author} {\bibfnamefont {D.}~\bibnamefont
  {Haug}}, \bibinfo {author} {\bibfnamefont {B.}~\bibnamefont {Keimer}},\ and\
  \bibinfo {author} {\bibfnamefont {A.}~\bibnamefont {Cavalleri}},\ }\href
  {https://doi.org/10.1103/PhysRevB.89.184516} {\bibfield  {journal} {\bibinfo
  {journal} {Phys. Rev. B}\ }\textbf {\bibinfo {volume} {89}},\ \bibinfo
  {pages} {184516} (\bibinfo {year} {2014})}\BibitemShut {NoStop}%
\bibitem [{\citenamefont {Hu}\ \emph {et~al.}(2014)\citenamefont {Hu},
  \citenamefont {Kaiser}, \citenamefont {Nicoletti}, \citenamefont {Hunt},
  \citenamefont {Gierz}, \citenamefont {Ho}, \citenamefont {Tacon},
  \citenamefont {Loew}, \citenamefont {Keimer},\ and\ \citenamefont
  {Cavalleri}}]{HuW2014}%
  \BibitemOpen
  \bibfield  {author} {\bibinfo {author} {\bibfnamefont {W.}~\bibnamefont
  {Hu}}, \bibinfo {author} {\bibfnamefont {S.}~\bibnamefont {Kaiser}}, \bibinfo
  {author} {\bibfnamefont {D.}~\bibnamefont {Nicoletti}}, \bibinfo {author}
  {\bibfnamefont {C.~R.}\ \bibnamefont {Hunt}}, \bibinfo {author}
  {\bibfnamefont {I.}~\bibnamefont {Gierz}}, \bibinfo {author} {\bibfnamefont
  {M.~C.}\ \bibnamefont {Ho}}, \bibinfo {author} {\bibfnamefont {M.~L.}\
  \bibnamefont {Tacon}}, \bibinfo {author} {\bibfnamefont {T.}~\bibnamefont
  {Loew}}, \bibinfo {author} {\bibfnamefont {B.}~\bibnamefont {Keimer}},\ and\
  \bibinfo {author} {\bibfnamefont {A.}~\bibnamefont {Cavalleri}},\ }\href
  {https://doi.org/10.1038/NMAT3963} {\bibfield  {journal} {\bibinfo  {journal}
  {Nat. Mat.}\ }\textbf {\bibinfo {volume} {13}},\ \bibinfo {pages} {705}
  (\bibinfo {year} {2014})}\BibitemShut {NoStop}%
\bibitem [{\citenamefont {F{\"o}rst}\ \emph {et~al.}(2014)\citenamefont
  {F{\"o}rst}, \citenamefont {Frano}, \citenamefont {Kaiser}, \citenamefont
  {Mankowsky}, \citenamefont {Hunt}, \citenamefont {Turner}, \citenamefont
  {Dakovski}, \citenamefont {Minitti}, \citenamefont {Robinson}, \citenamefont
  {Loew}, \citenamefont {Le~Tacon}, \citenamefont {Keimer}, \citenamefont
  {Hill}, \citenamefont {Cavalleri},\ and\ \citenamefont {Dhesi}}]{Forst2014}%
  \BibitemOpen
  \bibfield  {author} {\bibinfo {author} {\bibfnamefont {M.}~\bibnamefont
  {F{\"o}rst}}, \bibinfo {author} {\bibfnamefont {A.}~\bibnamefont {Frano}},
  \bibinfo {author} {\bibfnamefont {S.}~\bibnamefont {Kaiser}}, \bibinfo
  {author} {\bibfnamefont {R.}~\bibnamefont {Mankowsky}}, \bibinfo {author}
  {\bibfnamefont {C.~R.}\ \bibnamefont {Hunt}}, \bibinfo {author}
  {\bibfnamefont {J.~J.}\ \bibnamefont {Turner}}, \bibinfo {author}
  {\bibfnamefont {G.~L.}\ \bibnamefont {Dakovski}}, \bibinfo {author}
  {\bibfnamefont {M.~P.}\ \bibnamefont {Minitti}}, \bibinfo {author}
  {\bibfnamefont {J.}~\bibnamefont {Robinson}}, \bibinfo {author}
  {\bibfnamefont {T.}~\bibnamefont {Loew}}, \bibinfo {author} {\bibfnamefont
  {M.}~\bibnamefont {Le~Tacon}}, \bibinfo {author} {\bibfnamefont
  {B.}~\bibnamefont {Keimer}}, \bibinfo {author} {\bibfnamefont {J.~P.}\
  \bibnamefont {Hill}}, \bibinfo {author} {\bibfnamefont {A.}~\bibnamefont
  {Cavalleri}},\ and\ \bibinfo {author} {\bibfnamefont {S.~S.}\ \bibnamefont
  {Dhesi}},\ }\href {https://doi.org/10.1103/PhysRevB.90.184514} {\bibfield
  {journal} {\bibinfo  {journal} {Phys. Rev. B}\ }\textbf {\bibinfo {volume}
  {90}},\ \bibinfo {pages} {184514} (\bibinfo {year} {2014})}\BibitemShut
  {NoStop}%
\bibitem [{\citenamefont {Mankowsky}\ \emph {et~al.}(2015)\citenamefont
  {Mankowsky}, \citenamefont {F\"orst}, \citenamefont {Loew}, \citenamefont
  {Porras}, \citenamefont {Keimer},\ and\ \citenamefont
  {Cavalleri}}]{Mankowsky2015}%
  \BibitemOpen
  \bibfield  {author} {\bibinfo {author} {\bibfnamefont {R.}~\bibnamefont
  {Mankowsky}}, \bibinfo {author} {\bibfnamefont {M.}~\bibnamefont {F\"orst}},
  \bibinfo {author} {\bibfnamefont {T.}~\bibnamefont {Loew}}, \bibinfo {author}
  {\bibfnamefont {J.}~\bibnamefont {Porras}}, \bibinfo {author} {\bibfnamefont
  {B.}~\bibnamefont {Keimer}},\ and\ \bibinfo {author} {\bibfnamefont
  {A.}~\bibnamefont {Cavalleri}},\ }\href
  {https://doi.org/10.1103/PhysRevB.91.094308} {\bibfield  {journal} {\bibinfo
  {journal} {Phys. Rev. B}\ }\textbf {\bibinfo {volume} {91}},\ \bibinfo
  {pages} {094308} (\bibinfo {year} {2015})}\BibitemShut {NoStop}%
\bibitem [{\citenamefont {Mankowsky}\ \emph {et~al.}(2017)\citenamefont
  {Mankowsky}, \citenamefont {Fechner}, \citenamefont {Forst}, \citenamefont
  {von Hoegen}, \citenamefont {Porras}, \citenamefont {Loew}, \citenamefont
  {Dakovski}, \citenamefont {Seaberg}, \citenamefont {Moller}, \citenamefont
  {Coslovich}, \citenamefont {Keimer}, \citenamefont {Dhesi},\ and\
  \citenamefont {Cavalleri}}]{Mankowsky2017}%
  \BibitemOpen
  \bibfield  {author} {\bibinfo {author} {\bibfnamefont {R.}~\bibnamefont
  {Mankowsky}}, \bibinfo {author} {\bibfnamefont {M.}~\bibnamefont {Fechner}},
  \bibinfo {author} {\bibfnamefont {M.}~\bibnamefont {Forst}}, \bibinfo
  {author} {\bibfnamefont {A.}~\bibnamefont {von Hoegen}}, \bibinfo {author}
  {\bibfnamefont {J.}~\bibnamefont {Porras}}, \bibinfo {author} {\bibfnamefont
  {T.}~\bibnamefont {Loew}}, \bibinfo {author} {\bibfnamefont {G.~L.}\
  \bibnamefont {Dakovski}}, \bibinfo {author} {\bibfnamefont {M.}~\bibnamefont
  {Seaberg}}, \bibinfo {author} {\bibfnamefont {S.}~\bibnamefont {Moller}},
  \bibinfo {author} {\bibfnamefont {G.}~\bibnamefont {Coslovich}}, \bibinfo
  {author} {\bibfnamefont {B.}~\bibnamefont {Keimer}}, \bibinfo {author}
  {\bibfnamefont {S.~S.}\ \bibnamefont {Dhesi}},\ and\ \bibinfo {author}
  {\bibfnamefont {A.}~\bibnamefont {Cavalleri}},\ }\href
  {https://doi.org/10.1063/1.4977672} {\bibfield  {journal} {\bibinfo
  {journal} {Structural Dynamics}\ }\textbf {\bibinfo {volume} {4}},\ \bibinfo
  {pages} {044007} (\bibinfo {year} {2017})}\BibitemShut {NoStop}%
\bibitem [{\citenamefont {Mitrano}\ \emph {et~al.}(2016)\citenamefont
  {Mitrano}, \citenamefont {Cantaluppi}, \citenamefont {Nicoletti},
  \citenamefont {Kaiser}, \citenamefont {Perucchi}, \citenamefont {Lupi},
  \citenamefont {Pietro}, \citenamefont {Pontiroli}, \citenamefont {Ricc{\`o}},
  \citenamefont {Clark}, \citenamefont {Jaksch},\ and\ \citenamefont
  {Cavalleri}}]{Mitrano2016}%
  \BibitemOpen
  \bibfield  {author} {\bibinfo {author} {\bibfnamefont {M.}~\bibnamefont
  {Mitrano}}, \bibinfo {author} {\bibfnamefont {A.}~\bibnamefont {Cantaluppi}},
  \bibinfo {author} {\bibfnamefont {D.}~\bibnamefont {Nicoletti}}, \bibinfo
  {author} {\bibfnamefont {S.}~\bibnamefont {Kaiser}}, \bibinfo {author}
  {\bibfnamefont {A.}~\bibnamefont {Perucchi}}, \bibinfo {author}
  {\bibfnamefont {S.}~\bibnamefont {Lupi}}, \bibinfo {author} {\bibfnamefont
  {P.~D.}\ \bibnamefont {Pietro}}, \bibinfo {author} {\bibfnamefont
  {D.}~\bibnamefont {Pontiroli}}, \bibinfo {author} {\bibfnamefont
  {M.}~\bibnamefont {Ricc{\`o}}}, \bibinfo {author} {\bibfnamefont {S.~R.}\
  \bibnamefont {Clark}}, \bibinfo {author} {\bibfnamefont {D.}~\bibnamefont
  {Jaksch}},\ and\ \bibinfo {author} {\bibfnamefont {A.}~\bibnamefont
  {Cavalleri}},\ }\href {https://doi.org/10.1038/nature16522} {\bibfield
  {journal} {\bibinfo  {journal} {Nature}\ }\textbf {\bibinfo {volume} {530}},\
  \bibinfo {pages} {461} (\bibinfo {year} {2016})}\BibitemShut {NoStop}%
\bibitem [{\citenamefont {Cantaluppi}\ \emph {et~al.}(2018)\citenamefont
  {Cantaluppi}, \citenamefont {Buzzi}, \citenamefont {Jotzu}, \citenamefont
  {Nicoletti}, \citenamefont {Mitrano}, \citenamefont {Pontiroli},
  \citenamefont {Ricc{\`o}}, \citenamefont {Perucchi}, \citenamefont
  {Di~Pietro},\ and\ \citenamefont {Cavalleri}}]{Cantaluppi2018}%
  \BibitemOpen
  \bibfield  {author} {\bibinfo {author} {\bibfnamefont {A.}~\bibnamefont
  {Cantaluppi}}, \bibinfo {author} {\bibfnamefont {M.}~\bibnamefont {Buzzi}},
  \bibinfo {author} {\bibfnamefont {G.}~\bibnamefont {Jotzu}}, \bibinfo
  {author} {\bibfnamefont {D.}~\bibnamefont {Nicoletti}}, \bibinfo {author}
  {\bibfnamefont {M.}~\bibnamefont {Mitrano}}, \bibinfo {author} {\bibfnamefont
  {D.}~\bibnamefont {Pontiroli}}, \bibinfo {author} {\bibfnamefont
  {M.}~\bibnamefont {Ricc{\`o}}}, \bibinfo {author} {\bibfnamefont
  {A.}~\bibnamefont {Perucchi}}, \bibinfo {author} {\bibfnamefont
  {P.}~\bibnamefont {Di~Pietro}},\ and\ \bibinfo {author} {\bibfnamefont
  {A.}~\bibnamefont {Cavalleri}},\ }\href
  {https://doi.org/10.1038/s41567-018-0134-8} {\bibfield  {journal} {\bibinfo
  {journal} {Nat. Phys.}\ }\textbf {\bibinfo {volume} {14}},\ \bibinfo {pages}
  {837} (\bibinfo {year} {2018})}\BibitemShut {NoStop}%
\bibitem [{\citenamefont {Knap}\ \emph {et~al.}(2016)\citenamefont {Knap},
  \citenamefont {Babadi}, \citenamefont {Refael}, \citenamefont {Martin},\ and\
  \citenamefont {Demler}}]{Knap2016}%
  \BibitemOpen
  \bibfield  {author} {\bibinfo {author} {\bibfnamefont {M.}~\bibnamefont
  {Knap}}, \bibinfo {author} {\bibfnamefont {M.}~\bibnamefont {Babadi}},
  \bibinfo {author} {\bibfnamefont {G.}~\bibnamefont {Refael}}, \bibinfo
  {author} {\bibfnamefont {I.}~\bibnamefont {Martin}},\ and\ \bibinfo {author}
  {\bibfnamefont {E.}~\bibnamefont {Demler}},\ }\href
  {https://doi.org/10.1103/PhysRevB.94.214504} {\bibfield  {journal} {\bibinfo
  {journal} {Phys. Rev. B}\ }\textbf {\bibinfo {volume} {94}},\ \bibinfo
  {pages} {214504} (\bibinfo {year} {2016})}\BibitemShut {NoStop}%
\bibitem [{\citenamefont {Babadi}\ \emph {et~al.}(2017)\citenamefont {Babadi},
  \citenamefont {Knap}, \citenamefont {Martin}, \citenamefont {Refael},\ and\
  \citenamefont {Demler}}]{Babadi2017}%
  \BibitemOpen
  \bibfield  {author} {\bibinfo {author} {\bibfnamefont {M.}~\bibnamefont
  {Babadi}}, \bibinfo {author} {\bibfnamefont {M.}~\bibnamefont {Knap}},
  \bibinfo {author} {\bibfnamefont {I.}~\bibnamefont {Martin}}, \bibinfo
  {author} {\bibfnamefont {G.}~\bibnamefont {Refael}},\ and\ \bibinfo {author}
  {\bibfnamefont {E.}~\bibnamefont {Demler}},\ }\href
  {https://doi.org/10.1103/PhysRevB.96.014512} {\bibfield  {journal} {\bibinfo
  {journal} {Phys. Rev. B}\ }\textbf {\bibinfo {volume} {96}},\ \bibinfo
  {pages} {014512} (\bibinfo {year} {2017})}\BibitemShut {NoStop}%
\bibitem [{\citenamefont {Kennes}\ \emph {et~al.}(2017)\citenamefont {Kennes},
  \citenamefont {Wilner}, \citenamefont {Reichman},\ and\ \citenamefont
  {Millis}}]{Kennes2017}%
  \BibitemOpen
  \bibfield  {author} {\bibinfo {author} {\bibfnamefont {D.~M.}\ \bibnamefont
  {Kennes}}, \bibinfo {author} {\bibfnamefont {E.~Y.}\ \bibnamefont {Wilner}},
  \bibinfo {author} {\bibfnamefont {D.~R.}\ \bibnamefont {Reichman}},\ and\
  \bibinfo {author} {\bibfnamefont {A.~J.}\ \bibnamefont {Millis}},\ }\href
  {https://doi.org/10.1038/nphys4024} {\bibfield  {journal} {\bibinfo
  {journal} {Nat. Phys.}\ }\textbf {\bibinfo {volume} {13}},\ \bibinfo {pages}
  {479} (\bibinfo {year} {2017})}\BibitemShut {NoStop}%
\bibitem [{\citenamefont {Murakami}\ \emph {et~al.}(2017)\citenamefont
  {Murakami}, \citenamefont {Tsuji}, \citenamefont {Eckstein},\ and\
  \citenamefont {Werner}}]{Murakami2017}%
  \BibitemOpen
  \bibfield  {author} {\bibinfo {author} {\bibfnamefont {Y.}~\bibnamefont
  {Murakami}}, \bibinfo {author} {\bibfnamefont {N.}~\bibnamefont {Tsuji}},
  \bibinfo {author} {\bibfnamefont {M.}~\bibnamefont {Eckstein}},\ and\
  \bibinfo {author} {\bibfnamefont {P.}~\bibnamefont {Werner}},\ }\href
  {https://doi.org/10.1103/PhysRevB.96.045125} {\bibfield  {journal} {\bibinfo
  {journal} {Phys. Rev. B}\ }\textbf {\bibinfo {volume} {96}},\ \bibinfo
  {pages} {045125} (\bibinfo {year} {2017})}\BibitemShut {NoStop}%
\bibitem [{\citenamefont {Sentef}\ \emph {et~al.}(2017)\citenamefont {Sentef},
  \citenamefont {Tokuno}, \citenamefont {Georges},\ and\ \citenamefont
  {Kollath}}]{sentef2017}%
  \BibitemOpen
  \bibfield  {author} {\bibinfo {author} {\bibfnamefont {M.~A.}\ \bibnamefont
  {Sentef}}, \bibinfo {author} {\bibfnamefont {A.}~\bibnamefont {Tokuno}},
  \bibinfo {author} {\bibfnamefont {A.}~\bibnamefont {Georges}},\ and\ \bibinfo
  {author} {\bibfnamefont {C.}~\bibnamefont {Kollath}},\ }\href
  {https://doi.org/10.1103/PhysRevLett.118.087002} {\bibfield  {journal}
  {\bibinfo  {journal} {Phys. Rev. Lett.}\ }\textbf {\bibinfo {volume} {118}},\
  \bibinfo {pages} {087002} (\bibinfo {year} {2017})}\BibitemShut {NoStop}%
\bibitem [{\citenamefont {Patel}\ and\ \citenamefont
  {Eberlein}(2016)}]{patel2016}%
  \BibitemOpen
  \bibfield  {author} {\bibinfo {author} {\bibfnamefont {A.~A.}\ \bibnamefont
  {Patel}}\ and\ \bibinfo {author} {\bibfnamefont {A.}~\bibnamefont
  {Eberlein}},\ }\href {https://doi.org/10.1103/PhysRevB.93.195139} {\bibfield
  {journal} {\bibinfo  {journal} {Phys. Rev. B}\ }\textbf {\bibinfo {volume}
  {93}},\ \bibinfo {pages} {195139} (\bibinfo {year} {2016})}\BibitemShut
  {NoStop}%
\bibitem [{\citenamefont {Sentef}\ \emph {et~al.}(2016)\citenamefont {Sentef},
  \citenamefont {Kemper}, \citenamefont {Georges},\ and\ \citenamefont
  {Kollath}}]{sentef2016}%
  \BibitemOpen
  \bibfield  {author} {\bibinfo {author} {\bibfnamefont {M.~A.}\ \bibnamefont
  {Sentef}}, \bibinfo {author} {\bibfnamefont {A.~F.}\ \bibnamefont {Kemper}},
  \bibinfo {author} {\bibfnamefont {A.}~\bibnamefont {Georges}},\ and\ \bibinfo
  {author} {\bibfnamefont {C.}~\bibnamefont {Kollath}},\ }\href
  {https://doi.org/10.1103/PhysRevB.93.144506} {\bibfield  {journal} {\bibinfo
  {journal} {Phys. Rev. B}\ }\textbf {\bibinfo {volume} {93}},\ \bibinfo
  {pages} {144506} (\bibinfo {year} {2016})}\BibitemShut {NoStop}%
\bibitem [{\citenamefont {Ido}\ \emph {et~al.}(2017)\citenamefont {Ido},
  \citenamefont {Ohgoe},\ and\ \citenamefont {Imada}}]{ido2017}%
  \BibitemOpen
  \bibfield  {author} {\bibinfo {author} {\bibfnamefont {K.}~\bibnamefont
  {Ido}}, \bibinfo {author} {\bibfnamefont {T.}~\bibnamefont {Ohgoe}},\ and\
  \bibinfo {author} {\bibfnamefont {M.}~\bibnamefont {Imada}},\ }\href
  {https://doi.org/10.1126/sciadv.1700718} {\bibfield  {journal} {\bibinfo
  {journal} {Sci. Adv.}\ }\textbf {\bibinfo {volume} {3}},\ \bibinfo {pages}
  {e1700718} (\bibinfo {year} {2017})}\BibitemShut {NoStop}%
\bibitem [{\citenamefont {Mazza}\ and\ \citenamefont
  {Georges}(2017)}]{mazza2017}%
  \BibitemOpen
  \bibfield  {author} {\bibinfo {author} {\bibfnamefont {G.}~\bibnamefont
  {Mazza}}\ and\ \bibinfo {author} {\bibfnamefont {A.}~\bibnamefont
  {Georges}},\ }\href {https://doi.org/10.1103/PhysRevB.96.064515} {\bibfield
  {journal} {\bibinfo  {journal} {Phys. Rev. B}\ }\textbf {\bibinfo {volume}
  {96}},\ \bibinfo {pages} {064515} (\bibinfo {year} {2017})}\BibitemShut
  {NoStop}%
\bibitem [{\citenamefont {Wang}\ \emph {et~al.}(2018)\citenamefont {Wang},
  \citenamefont {Chen}, \citenamefont {Moritz},\ and\ \citenamefont
  {Devereaux}}]{wang2018}%
  \BibitemOpen
  \bibfield  {author} {\bibinfo {author} {\bibfnamefont {Y.}~\bibnamefont
  {Wang}}, \bibinfo {author} {\bibfnamefont {C.-C.}\ \bibnamefont {Chen}},
  \bibinfo {author} {\bibfnamefont {B.}~\bibnamefont {Moritz}},\ and\ \bibinfo
  {author} {\bibfnamefont {T.~P.}\ \bibnamefont {Devereaux}},\ }\href
  {https://doi.org/10.1103/PhysRevLett.120.246402} {\bibfield  {journal}
  {\bibinfo  {journal} {Phys. Rev. Lett.}\ }\textbf {\bibinfo {volume} {120}},\
  \bibinfo {pages} {246402} (\bibinfo {year} {2018})}\BibitemShut {NoStop}%
\bibitem [{\citenamefont {Bittner}\ \emph {et~al.}(2019)\citenamefont
  {Bittner}, \citenamefont {Tohyama}, \citenamefont {Kaiser},\ and\
  \citenamefont {Manske}}]{bittner2019}%
  \BibitemOpen
  \bibfield  {author} {\bibinfo {author} {\bibfnamefont {N.}~\bibnamefont
  {Bittner}}, \bibinfo {author} {\bibfnamefont {T.}~\bibnamefont {Tohyama}},
  \bibinfo {author} {\bibfnamefont {S.}~\bibnamefont {Kaiser}},\ and\ \bibinfo
  {author} {\bibfnamefont {D.}~\bibnamefont {Manske}},\ }\href
  {https://doi.org/10.7566/JPSJ.88.044704} {\bibfield  {journal} {\bibinfo
  {journal} {J. Phys. Soc. Jpn.}\ }\textbf {\bibinfo {volume} {88}},\ \bibinfo
  {pages} {044704} (\bibinfo {year} {2019})}\BibitemShut {NoStop}%
\bibitem [{\citenamefont {Kaneko}\ \emph {et~al.}(2020)\citenamefont {Kaneko},
  \citenamefont {Yunoki},\ and\ \citenamefont {Millis}}]{kaneko2020}%
  \BibitemOpen
  \bibfield  {author} {\bibinfo {author} {\bibfnamefont {T.}~\bibnamefont
  {Kaneko}}, \bibinfo {author} {\bibfnamefont {S.}~\bibnamefont {Yunoki}},\
  and\ \bibinfo {author} {\bibfnamefont {A.~J.}\ \bibnamefont {Millis}},\
  }\href {https://doi.org/10.1103/PhysRevResearch.2.032027} {\bibfield
  {journal} {\bibinfo  {journal} {Phys. Rev. Research}\ }\textbf {\bibinfo
  {volume} {2}},\ \bibinfo {pages} {032027} (\bibinfo {year}
  {2020})}\BibitemShut {NoStop}%
\bibitem [{\citenamefont {Kaneko}\ \emph {et~al.}(2019)\citenamefont {Kaneko},
  \citenamefont {Shirakawa}, \citenamefont {Sorella},\ and\ \citenamefont
  {Yunoki}}]{kaneko2019a}%
  \BibitemOpen
  \bibfield  {author} {\bibinfo {author} {\bibfnamefont {T.}~\bibnamefont
  {Kaneko}}, \bibinfo {author} {\bibfnamefont {T.}~\bibnamefont {Shirakawa}},
  \bibinfo {author} {\bibfnamefont {S.}~\bibnamefont {Sorella}},\ and\ \bibinfo
  {author} {\bibfnamefont {S.}~\bibnamefont {Yunoki}},\ }\href
  {https://doi.org/10.1103/PhysRevLett.122.077002} {\bibfield  {journal}
  {\bibinfo  {journal} {Phys. Rev. Lett.}\ }\textbf {\bibinfo {volume} {122}},\
  \bibinfo {pages} {077002} (\bibinfo {year} {2019})}\BibitemShut {NoStop}%
\bibitem [{\citenamefont {Li}\ \emph {et~al.}(2019)\citenamefont {Li},
  \citenamefont {Golez}, \citenamefont {Werner},\ and\ \citenamefont
  {Eckstein}}]{li2019a}%
  \BibitemOpen
  \bibfield  {author} {\bibinfo {author} {\bibfnamefont {J.}~\bibnamefont
  {Li}}, \bibinfo {author} {\bibfnamefont {D.}~\bibnamefont {Golez}}, \bibinfo
  {author} {\bibfnamefont {P.}~\bibnamefont {Werner}},\ and\ \bibinfo {author}
  {\bibfnamefont {M.}~\bibnamefont {Eckstein}},\ }\href@noop {} {\  (\bibinfo
  {year} {2019})},\ \Eprint {https://arxiv.org/abs/arXiv:1908.08693}
  {arXiv:1908.08693} \BibitemShut {NoStop}%
\bibitem [{\citenamefont {Nava}\ \emph {et~al.}(2018)\citenamefont {Nava},
  \citenamefont {Giannetti}, \citenamefont {Georges}, \citenamefont {Tosatti},\
  and\ \citenamefont {Fabrizio}}]{nava2018}%
  \BibitemOpen
  \bibfield  {author} {\bibinfo {author} {\bibfnamefont {A.}~\bibnamefont
  {Nava}}, \bibinfo {author} {\bibfnamefont {C.}~\bibnamefont {Giannetti}},
  \bibinfo {author} {\bibfnamefont {A.}~\bibnamefont {Georges}}, \bibinfo
  {author} {\bibfnamefont {E.}~\bibnamefont {Tosatti}},\ and\ \bibinfo {author}
  {\bibfnamefont {M.}~\bibnamefont {Fabrizio}},\ }\href
  {https://doi.org/10.1038/nphys4288} {\bibfield  {journal} {\bibinfo
  {journal} {Nat. Phys.}\ }\textbf {\bibinfo {volume} {14}},\ \bibinfo {pages}
  {154} (\bibinfo {year} {2018})}\BibitemShut {NoStop}%
\bibitem [{\citenamefont {Werner}\ \emph {et~al.}(2018)\citenamefont {Werner},
  \citenamefont {Strand}, \citenamefont {Hoshino}, \citenamefont {Murakami},\
  and\ \citenamefont {Eckstein}}]{werner2018}%
  \BibitemOpen
  \bibfield  {author} {\bibinfo {author} {\bibfnamefont {P.}~\bibnamefont
  {Werner}}, \bibinfo {author} {\bibfnamefont {H.~U.~R.}\ \bibnamefont
  {Strand}}, \bibinfo {author} {\bibfnamefont {S.}~\bibnamefont {Hoshino}},
  \bibinfo {author} {\bibfnamefont {Y.}~\bibnamefont {Murakami}},\ and\
  \bibinfo {author} {\bibfnamefont {M.}~\bibnamefont {Eckstein}},\ }\href
  {https://doi.org/10.1103/PhysRevB.97.165119} {\bibfield  {journal} {\bibinfo
  {journal} {Phys. Rev. B}\ }\textbf {\bibinfo {volume} {97}},\ \bibinfo
  {pages} {165119} (\bibinfo {year} {2018})}\BibitemShut {NoStop}%
\bibitem [{\citenamefont {Denny}\ \emph {et~al.}(2015)\citenamefont {Denny},
  \citenamefont {Clark}, \citenamefont {Laplace}, \citenamefont {Cavalleri},\
  and\ \citenamefont {Jaksch}}]{denny2015}%
  \BibitemOpen
  \bibfield  {author} {\bibinfo {author} {\bibfnamefont {S.~J.}\ \bibnamefont
  {Denny}}, \bibinfo {author} {\bibfnamefont {S.~R.}\ \bibnamefont {Clark}},
  \bibinfo {author} {\bibfnamefont {Y.}~\bibnamefont {Laplace}}, \bibinfo
  {author} {\bibfnamefont {A.}~\bibnamefont {Cavalleri}},\ and\ \bibinfo
  {author} {\bibfnamefont {D.}~\bibnamefont {Jaksch}},\ }\href
  {https://doi.org/10.1103/PhysRevLett.114.137001} {\bibfield  {journal}
  {\bibinfo  {journal} {Phys. Rev. Lett.}\ }\textbf {\bibinfo {volume} {114}},\
  \bibinfo {pages} {137001} (\bibinfo {year} {2015})}\BibitemShut {NoStop}%
\bibitem [{\citenamefont {H{\"o}ppner}\ \emph {et~al.}(2015)\citenamefont
  {H{\"o}ppner}, \citenamefont {Zhu}, \citenamefont {Rexin}, \citenamefont
  {Cavalleri},\ and\ \citenamefont {Mathey}}]{hoppner2015}%
  \BibitemOpen
  \bibfield  {author} {\bibinfo {author} {\bibfnamefont {R.}~\bibnamefont
  {H{\"o}ppner}}, \bibinfo {author} {\bibfnamefont {B.}~\bibnamefont {Zhu}},
  \bibinfo {author} {\bibfnamefont {T.}~\bibnamefont {Rexin}}, \bibinfo
  {author} {\bibfnamefont {A.}~\bibnamefont {Cavalleri}},\ and\ \bibinfo
  {author} {\bibfnamefont {L.}~\bibnamefont {Mathey}},\ }\href
  {https://doi.org/10.1103/PhysRevB.91.104507} {\bibfield  {journal} {\bibinfo
  {journal} {Phys. Rev. B}\ }\textbf {\bibinfo {volume} {91}},\ \bibinfo
  {pages} {104507} (\bibinfo {year} {2015})}\BibitemShut {NoStop}%
\bibitem [{\citenamefont {Okamoto}\ \emph {et~al.}(2016)\citenamefont
  {Okamoto}, \citenamefont {Cavalleri},\ and\ \citenamefont
  {Mathey}}]{okamoto2016}%
  \BibitemOpen
  \bibfield  {author} {\bibinfo {author} {\bibfnamefont {J.-i.}\ \bibnamefont
  {Okamoto}}, \bibinfo {author} {\bibfnamefont {A.}~\bibnamefont {Cavalleri}},\
  and\ \bibinfo {author} {\bibfnamefont {L.}~\bibnamefont {Mathey}},\ }\href
  {https://doi.org/10.1103/PhysRevLett.117.227001} {\bibfield  {journal}
  {\bibinfo  {journal} {Phys. Rev. Lett.}\ }\textbf {\bibinfo {volume} {117}},\
  \bibinfo {pages} {227001} (\bibinfo {year} {2016})}\BibitemShut {NoStop}%
\bibitem [{\citenamefont {Okamoto}\ \emph {et~al.}(2017)\citenamefont
  {Okamoto}, \citenamefont {Hu}, \citenamefont {Cavalleri},\ and\ \citenamefont
  {Mathey}}]{okamoto2017}%
  \BibitemOpen
  \bibfield  {author} {\bibinfo {author} {\bibfnamefont {J.-i.}\ \bibnamefont
  {Okamoto}}, \bibinfo {author} {\bibfnamefont {W.}~\bibnamefont {Hu}},
  \bibinfo {author} {\bibfnamefont {A.}~\bibnamefont {Cavalleri}},\ and\
  \bibinfo {author} {\bibfnamefont {L.}~\bibnamefont {Mathey}},\ }\href
  {https://doi.org/10.1103/PhysRevB.96.144505} {\bibfield  {journal} {\bibinfo
  {journal} {Phys. Rev. B}\ }\textbf {\bibinfo {volume} {96}},\ \bibinfo
  {pages} {144505} (\bibinfo {year} {2017})}\BibitemShut {NoStop}%
\bibitem [{\citenamefont {Schlawin}\ \emph {et~al.}(2017)\citenamefont
  {Schlawin}, \citenamefont {Dietrich}, \citenamefont {Kiffner}, \citenamefont
  {Cavalleri},\ and\ \citenamefont {Jaksch}}]{schlawin2017}%
  \BibitemOpen
  \bibfield  {author} {\bibinfo {author} {\bibfnamefont {F.}~\bibnamefont
  {Schlawin}}, \bibinfo {author} {\bibfnamefont {A.~S.~D.}\ \bibnamefont
  {Dietrich}}, \bibinfo {author} {\bibfnamefont {M.}~\bibnamefont {Kiffner}},
  \bibinfo {author} {\bibfnamefont {A.}~\bibnamefont {Cavalleri}},\ and\
  \bibinfo {author} {\bibfnamefont {D.}~\bibnamefont {Jaksch}},\ }\href
  {https://doi.org/10.1103/PhysRevB.96.064526} {\bibfield  {journal} {\bibinfo
  {journal} {Phys. Rev. B}\ }\textbf {\bibinfo {volume} {96}},\ \bibinfo
  {pages} {064526} (\bibinfo {year} {2017})}\BibitemShut {NoStop}%
\bibitem [{\citenamefont {Chiriac{\`o}}\ \emph {et~al.}(2018)\citenamefont
  {Chiriac{\`o}}, \citenamefont {Millis},\ and\ \citenamefont
  {Aleiner}}]{chiriaco2018}%
  \BibitemOpen
  \bibfield  {author} {\bibinfo {author} {\bibfnamefont {G.}~\bibnamefont
  {Chiriac{\`o}}}, \bibinfo {author} {\bibfnamefont {A.~J.}\ \bibnamefont
  {Millis}},\ and\ \bibinfo {author} {\bibfnamefont {I.~L.}\ \bibnamefont
  {Aleiner}},\ }\href {https://doi.org/10.1103/PhysRevB.98.220510} {\bibfield
  {journal} {\bibinfo  {journal} {Phys. Rev. B}\ }\textbf {\bibinfo {volume}
  {98}},\ \bibinfo {pages} {220510} (\bibinfo {year} {2018})}\BibitemShut
  {NoStop}%
\bibitem [{\citenamefont {Harland}\ \emph {et~al.}(2019)\citenamefont
  {Harland}, \citenamefont {Brener}, \citenamefont {Lichtenstein},\ and\
  \citenamefont {Katsnelson}}]{harland2019}%
  \BibitemOpen
  \bibfield  {author} {\bibinfo {author} {\bibfnamefont {M.}~\bibnamefont
  {Harland}}, \bibinfo {author} {\bibfnamefont {S.}~\bibnamefont {Brener}},
  \bibinfo {author} {\bibfnamefont {A.~I.}\ \bibnamefont {Lichtenstein}},\ and\
  \bibinfo {author} {\bibfnamefont {M.~I.}\ \bibnamefont {Katsnelson}},\ }\href
  {https://doi.org/10.1103/PhysRevB.100.024510} {\bibfield  {journal} {\bibinfo
   {journal} {Phys. Rev. B}\ }\textbf {\bibinfo {volume} {100}},\ \bibinfo
  {pages} {024510} (\bibinfo {year} {2019})}\BibitemShut {NoStop}%
\bibitem [{\citenamefont {Iwazaki}\ \emph {et~al.}(2019)\citenamefont
  {Iwazaki}, \citenamefont {Tsuji},\ and\ \citenamefont
  {Hoshino}}]{iwazaki2019}%
  \BibitemOpen
  \bibfield  {author} {\bibinfo {author} {\bibfnamefont {R.}~\bibnamefont
  {Iwazaki}}, \bibinfo {author} {\bibfnamefont {N.}~\bibnamefont {Tsuji}},\
  and\ \bibinfo {author} {\bibfnamefont {S.}~\bibnamefont {Hoshino}},\ }\href
  {https://doi.org/10.1103/PhysRevB.100.104521} {\bibfield  {journal} {\bibinfo
   {journal} {Phys. Rev. B}\ }\textbf {\bibinfo {volume} {100}},\ \bibinfo
  {pages} {104521} (\bibinfo {year} {2019})}\BibitemShut {NoStop}%
\bibitem [{\citenamefont {Lemonik}\ and\ \citenamefont
  {Mitra}(2019)}]{lemonik2019}%
  \BibitemOpen
  \bibfield  {author} {\bibinfo {author} {\bibfnamefont {Y.}~\bibnamefont
  {Lemonik}}\ and\ \bibinfo {author} {\bibfnamefont {A.}~\bibnamefont
  {Mitra}},\ }\href {https://doi.org/10.1103/PhysRevB.100.094503} {\bibfield
  {journal} {\bibinfo  {journal} {Phys. Rev. B}\ }\textbf {\bibinfo {volume}
  {100}},\ \bibinfo {pages} {094503} (\bibinfo {year} {2019})}\BibitemShut
  {NoStop}%
\bibitem [{\citenamefont {Cataliotti}\ \emph {et~al.}(2001)\citenamefont
  {Cataliotti}, \citenamefont {Burger}, \citenamefont {Fort}, \citenamefont
  {Maddaloni}, \citenamefont {Minardi}, \citenamefont {Trombettoni},
  \citenamefont {Smerzi},\ and\ \citenamefont {Inguscio}}]{Cataliotti2001}%
  \BibitemOpen
  \bibfield  {author} {\bibinfo {author} {\bibfnamefont {F.~S.}\ \bibnamefont
  {Cataliotti}}, \bibinfo {author} {\bibfnamefont {S.}~\bibnamefont {Burger}},
  \bibinfo {author} {\bibfnamefont {C.}~\bibnamefont {Fort}}, \bibinfo {author}
  {\bibfnamefont {P.}~\bibnamefont {Maddaloni}}, \bibinfo {author}
  {\bibfnamefont {F.}~\bibnamefont {Minardi}}, \bibinfo {author} {\bibfnamefont
  {A.}~\bibnamefont {Trombettoni}}, \bibinfo {author} {\bibfnamefont
  {A.}~\bibnamefont {Smerzi}},\ and\ \bibinfo {author} {\bibfnamefont
  {M.}~\bibnamefont {Inguscio}},\ }\href
  {https://doi.org/10.1126/science.1062612} {\bibfield  {journal} {\bibinfo
  {journal} {Science}\ }\textbf {\bibinfo {volume} {293}},\ \bibinfo {pages}
  {843} (\bibinfo {year} {2001})}\BibitemShut {NoStop}%
\bibitem [{\citenamefont {Albiez}\ \emph {et~al.}(2005)\citenamefont {Albiez},
  \citenamefont {Gati}, \citenamefont {F\"olling}, \citenamefont {Hunsmann},
  \citenamefont {Cristiani},\ and\ \citenamefont {Oberthaler}}]{Albiez2005}%
  \BibitemOpen
  \bibfield  {author} {\bibinfo {author} {\bibfnamefont {M.}~\bibnamefont
  {Albiez}}, \bibinfo {author} {\bibfnamefont {R.}~\bibnamefont {Gati}},
  \bibinfo {author} {\bibfnamefont {J.}~\bibnamefont {F\"olling}}, \bibinfo
  {author} {\bibfnamefont {S.}~\bibnamefont {Hunsmann}}, \bibinfo {author}
  {\bibfnamefont {M.}~\bibnamefont {Cristiani}},\ and\ \bibinfo {author}
  {\bibfnamefont {M.~K.}\ \bibnamefont {Oberthaler}},\ }\href
  {https://doi.org/10.1103/PhysRevLett.95.010402} {\bibfield  {journal}
  {\bibinfo  {journal} {Phys. Rev. Lett.}\ }\textbf {\bibinfo {volume} {95}},\
  \bibinfo {pages} {010402} (\bibinfo {year} {2005})}\BibitemShut {NoStop}%
\bibitem [{\citenamefont {Levy}\ \emph {et~al.}(2007)\citenamefont {Levy},
  \citenamefont {Lahoud}, \citenamefont {Shomroni},\ and\ \citenamefont
  {Steinhauer}}]{Levy2007}%
  \BibitemOpen
  \bibfield  {author} {\bibinfo {author} {\bibfnamefont {S.}~\bibnamefont
  {Levy}}, \bibinfo {author} {\bibfnamefont {E.}~\bibnamefont {Lahoud}},
  \bibinfo {author} {\bibfnamefont {I.}~\bibnamefont {Shomroni}},\ and\
  \bibinfo {author} {\bibfnamefont {J.}~\bibnamefont {Steinhauer}},\ }\href
  {https://doi.org/10.1038/nature06186} {\bibfield  {journal} {\bibinfo
  {journal} {Nature}\ }\textbf {\bibinfo {volume} {449}},\ \bibinfo {pages}
  {579} (\bibinfo {year} {2007})}\BibitemShut {NoStop}%
\bibitem [{\citenamefont {LeBlanc}\ \emph {et~al.}(2011)\citenamefont
  {LeBlanc}, \citenamefont {Bardon}, \citenamefont {McKeever}, \citenamefont
  {Extavour}, \citenamefont {Jervis}, \citenamefont {Thywissen}, \citenamefont
  {Piazza},\ and\ \citenamefont {Smerzi}}]{LeBlanc2011}%
  \BibitemOpen
  \bibfield  {author} {\bibinfo {author} {\bibfnamefont {L.~J.}\ \bibnamefont
  {LeBlanc}}, \bibinfo {author} {\bibfnamefont {A.~B.}\ \bibnamefont {Bardon}},
  \bibinfo {author} {\bibfnamefont {J.}~\bibnamefont {McKeever}}, \bibinfo
  {author} {\bibfnamefont {M.~H.~T.}\ \bibnamefont {Extavour}}, \bibinfo
  {author} {\bibfnamefont {D.}~\bibnamefont {Jervis}}, \bibinfo {author}
  {\bibfnamefont {J.~H.}\ \bibnamefont {Thywissen}}, \bibinfo {author}
  {\bibfnamefont {F.}~\bibnamefont {Piazza}},\ and\ \bibinfo {author}
  {\bibfnamefont {A.}~\bibnamefont {Smerzi}},\ }\href
  {https://doi.org/10.1103/PhysRevLett.106.025302} {\bibfield  {journal}
  {\bibinfo  {journal} {Phys. Rev. Lett.}\ }\textbf {\bibinfo {volume} {106}},\
  \bibinfo {pages} {025302} (\bibinfo {year} {2011})}\BibitemShut {NoStop}%
\bibitem [{\citenamefont {Betz}\ \emph {et~al.}(2011)\citenamefont {Betz},
  \citenamefont {Manz}, \citenamefont {B\"ucker}, \citenamefont {Berrada},
  \citenamefont {Koller}, \citenamefont {Kazakov}, \citenamefont {Mazets},
  \citenamefont {Stimming}, \citenamefont {Perrin}, \citenamefont {Schumm},\
  and\ \citenamefont {Schmiedmayer}}]{Betz2011}%
  \BibitemOpen
  \bibfield  {author} {\bibinfo {author} {\bibfnamefont {T.}~\bibnamefont
  {Betz}}, \bibinfo {author} {\bibfnamefont {S.}~\bibnamefont {Manz}}, \bibinfo
  {author} {\bibfnamefont {R.}~\bibnamefont {B\"ucker}}, \bibinfo {author}
  {\bibfnamefont {T.}~\bibnamefont {Berrada}}, \bibinfo {author} {\bibfnamefont
  {C.}~\bibnamefont {Koller}}, \bibinfo {author} {\bibfnamefont
  {G.}~\bibnamefont {Kazakov}}, \bibinfo {author} {\bibfnamefont {I.~E.}\
  \bibnamefont {Mazets}}, \bibinfo {author} {\bibfnamefont {H.-P.}\
  \bibnamefont {Stimming}}, \bibinfo {author} {\bibfnamefont {A.}~\bibnamefont
  {Perrin}}, \bibinfo {author} {\bibfnamefont {T.}~\bibnamefont {Schumm}},\
  and\ \bibinfo {author} {\bibfnamefont {J.}~\bibnamefont {Schmiedmayer}},\
  }\href {https://doi.org/10.1103/PhysRevLett.106.020407} {\bibfield  {journal}
  {\bibinfo  {journal} {Phys. Rev. Lett.}\ }\textbf {\bibinfo {volume} {106}},\
  \bibinfo {pages} {020407} (\bibinfo {year} {2011})}\BibitemShut {NoStop}%
\bibitem [{\citenamefont {Spagnolli}\ \emph {et~al.}(2017)\citenamefont
  {Spagnolli}, \citenamefont {Semeghini}, \citenamefont {Masi}, \citenamefont
  {Ferioli}, \citenamefont {Trenkwalder}, \citenamefont {Coop}, \citenamefont
  {Landini}, \citenamefont {Pezz\`e}, \citenamefont {Modugno}, \citenamefont
  {Inguscio}, \citenamefont {Smerzi},\ and\ \citenamefont
  {Fattori}}]{Spagnolli2017}%
  \BibitemOpen
  \bibfield  {author} {\bibinfo {author} {\bibfnamefont {G.}~\bibnamefont
  {Spagnolli}}, \bibinfo {author} {\bibfnamefont {G.}~\bibnamefont
  {Semeghini}}, \bibinfo {author} {\bibfnamefont {L.}~\bibnamefont {Masi}},
  \bibinfo {author} {\bibfnamefont {G.}~\bibnamefont {Ferioli}}, \bibinfo
  {author} {\bibfnamefont {A.}~\bibnamefont {Trenkwalder}}, \bibinfo {author}
  {\bibfnamefont {S.}~\bibnamefont {Coop}}, \bibinfo {author} {\bibfnamefont
  {M.}~\bibnamefont {Landini}}, \bibinfo {author} {\bibfnamefont
  {L.}~\bibnamefont {Pezz\`e}}, \bibinfo {author} {\bibfnamefont
  {G.}~\bibnamefont {Modugno}}, \bibinfo {author} {\bibfnamefont
  {M.}~\bibnamefont {Inguscio}}, \bibinfo {author} {\bibfnamefont
  {A.}~\bibnamefont {Smerzi}},\ and\ \bibinfo {author} {\bibfnamefont
  {M.}~\bibnamefont {Fattori}},\ }\href
  {https://doi.org/10.1103/PhysRevLett.118.230403} {\bibfield  {journal}
  {\bibinfo  {journal} {Phys. Rev. Lett.}\ }\textbf {\bibinfo {volume} {118}},\
  \bibinfo {pages} {230403} (\bibinfo {year} {2017})}\BibitemShut {NoStop}%
\bibitem [{\citenamefont {Valtolina}\ \emph {et~al.}(2015)\citenamefont
  {Valtolina}, \citenamefont {Burchianti}, \citenamefont {Amico}, \citenamefont
  {Neri}, \citenamefont {Xhani}, \citenamefont {Seman}, \citenamefont
  {Trombettoni}, \citenamefont {Smerzi}, \citenamefont {Zaccanti},
  \citenamefont {Inguscio},\ and\ \citenamefont {Roati}}]{Valtolina1505}%
  \BibitemOpen
  \bibfield  {author} {\bibinfo {author} {\bibfnamefont {G.}~\bibnamefont
  {Valtolina}}, \bibinfo {author} {\bibfnamefont {A.}~\bibnamefont
  {Burchianti}}, \bibinfo {author} {\bibfnamefont {A.}~\bibnamefont {Amico}},
  \bibinfo {author} {\bibfnamefont {E.}~\bibnamefont {Neri}}, \bibinfo {author}
  {\bibfnamefont {K.}~\bibnamefont {Xhani}}, \bibinfo {author} {\bibfnamefont
  {J.~A.}\ \bibnamefont {Seman}}, \bibinfo {author} {\bibfnamefont
  {A.}~\bibnamefont {Trombettoni}}, \bibinfo {author} {\bibfnamefont
  {A.}~\bibnamefont {Smerzi}}, \bibinfo {author} {\bibfnamefont
  {M.}~\bibnamefont {Zaccanti}}, \bibinfo {author} {\bibfnamefont
  {M.}~\bibnamefont {Inguscio}},\ and\ \bibinfo {author} {\bibfnamefont
  {G.}~\bibnamefont {Roati}},\ }\href {https://doi.org/10.1126/science.aac9725}
  {\bibfield  {journal} {\bibinfo  {journal} {Science}\ }\textbf {\bibinfo
  {volume} {350}},\ \bibinfo {pages} {1505} (\bibinfo {year}
  {2015})}\BibitemShut {NoStop}%
\bibitem [{\citenamefont {Chien}\ \emph {et~al.}(2015)\citenamefont {Chien},
  \citenamefont {Peotta},\ and\ \citenamefont {Di~Ventra}}]{Chien2015}%
  \BibitemOpen
  \bibfield  {author} {\bibinfo {author} {\bibfnamefont {C.-C.}\ \bibnamefont
  {Chien}}, \bibinfo {author} {\bibfnamefont {S.}~\bibnamefont {Peotta}},\ and\
  \bibinfo {author} {\bibfnamefont {M.}~\bibnamefont {Di~Ventra}},\ }\href
  {https://doi.org/10.1038/nphys3531} {\bibfield  {journal} {\bibinfo
  {journal} {Nat. Phys.}\ }\textbf {\bibinfo {volume} {11}},\ \bibinfo {pages}
  {998} (\bibinfo {year} {2015})}\BibitemShut {NoStop}%
\bibitem [{\citenamefont {Krinner}\ \emph {et~al.}(2017)\citenamefont
  {Krinner}, \citenamefont {Esslinger},\ and\ \citenamefont
  {Brantut}}]{Krinner2017}%
  \BibitemOpen
  \bibfield  {author} {\bibinfo {author} {\bibfnamefont {S.}~\bibnamefont
  {Krinner}}, \bibinfo {author} {\bibfnamefont {T.}~\bibnamefont {Esslinger}},\
  and\ \bibinfo {author} {\bibfnamefont {J.-P.}\ \bibnamefont {Brantut}},\
  }\href {https://doi.org/10.1088/1361-648x/aa74a1} {\bibfield  {journal}
  {\bibinfo  {journal} {Journal of Physics: Condensed Matter}\ }\textbf
  {\bibinfo {volume} {29}},\ \bibinfo {pages} {343003} (\bibinfo {year}
  {2017})}\BibitemShut {NoStop}%
\bibitem [{\citenamefont {Anderson}\ \emph {et~al.}(2019)\citenamefont
  {Anderson}, \citenamefont {Wang}, \citenamefont {Xu}, \citenamefont {Venu},
  \citenamefont {Trotzky}, \citenamefont {Chevy},\ and\ \citenamefont
  {Thywissen}}]{Anderson2019}%
  \BibitemOpen
  \bibfield  {author} {\bibinfo {author} {\bibfnamefont {R.}~\bibnamefont
  {Anderson}}, \bibinfo {author} {\bibfnamefont {F.}~\bibnamefont {Wang}},
  \bibinfo {author} {\bibfnamefont {P.}~\bibnamefont {Xu}}, \bibinfo {author}
  {\bibfnamefont {V.}~\bibnamefont {Venu}}, \bibinfo {author} {\bibfnamefont
  {S.}~\bibnamefont {Trotzky}}, \bibinfo {author} {\bibfnamefont
  {F.}~\bibnamefont {Chevy}},\ and\ \bibinfo {author} {\bibfnamefont {J.~H.}\
  \bibnamefont {Thywissen}},\ }\href
  {https://doi.org/10.1103/PhysRevLett.122.153602} {\bibfield  {journal}
  {\bibinfo  {journal} {Phys. Rev. Lett.}\ }\textbf {\bibinfo {volume} {122}},\
  \bibinfo {pages} {153602} (\bibinfo {year} {2019})}\BibitemShut {NoStop}%
\bibitem [{\citenamefont {Luick}\ \emph {et~al.}(2020)\citenamefont {Luick},
  \citenamefont {Sobirey}, \citenamefont {Bohlen}, \citenamefont {Singh},
  \citenamefont {Mathey}, \citenamefont {Lompe},\ and\ \citenamefont
  {Moritz}}]{Luick2020}%
  \BibitemOpen
  \bibfield  {author} {\bibinfo {author} {\bibfnamefont {N.}~\bibnamefont
  {Luick}}, \bibinfo {author} {\bibfnamefont {L.}~\bibnamefont {Sobirey}},
  \bibinfo {author} {\bibfnamefont {M.}~\bibnamefont {Bohlen}}, \bibinfo
  {author} {\bibfnamefont {V.~P.}\ \bibnamefont {Singh}}, \bibinfo {author}
  {\bibfnamefont {L.}~\bibnamefont {Mathey}}, \bibinfo {author} {\bibfnamefont
  {T.}~\bibnamefont {Lompe}},\ and\ \bibinfo {author} {\bibfnamefont
  {H.}~\bibnamefont {Moritz}},\ }\href
  {https://doi.org/10.1126/science.aaz2342} {\bibfield  {journal} {\bibinfo
  {journal} {Science}\ }\textbf {\bibinfo {volume} {369}},\ \bibinfo {pages}
  {89} (\bibinfo {year} {2020})}\BibitemShut {NoStop}%
\bibitem [{\citenamefont {Singh}\ \emph {et~al.}(2020)\citenamefont {Singh},
  \citenamefont {Luick}, \citenamefont {Sobirey},\ and\ \citenamefont
  {Mathey}}]{Singh2020}%
  \BibitemOpen
  \bibfield  {author} {\bibinfo {author} {\bibfnamefont {V.~P.}\ \bibnamefont
  {Singh}}, \bibinfo {author} {\bibfnamefont {N.}~\bibnamefont {Luick}},
  \bibinfo {author} {\bibfnamefont {L.}~\bibnamefont {Sobirey}},\ and\ \bibinfo
  {author} {\bibfnamefont {L.}~\bibnamefont {Mathey}},\ }\href
  {https://doi.org/10.1103/PhysRevResearch.2.033298} {\bibfield  {journal}
  {\bibinfo  {journal} {Phys. Rev. Research}\ }\textbf {\bibinfo {volume}
  {2}},\ \bibinfo {pages} {033298} (\bibinfo {year} {2020})}\BibitemShut
  {NoStop}%
\bibitem [{\citenamefont {Smerzi}\ \emph {et~al.}(1997)\citenamefont {Smerzi},
  \citenamefont {Fantoni}, \citenamefont {Giovanazzi},\ and\ \citenamefont
  {Shenoy}}]{Smerzi1997}%
  \BibitemOpen
  \bibfield  {author} {\bibinfo {author} {\bibfnamefont {A.}~\bibnamefont
  {Smerzi}}, \bibinfo {author} {\bibfnamefont {S.}~\bibnamefont {Fantoni}},
  \bibinfo {author} {\bibfnamefont {S.}~\bibnamefont {Giovanazzi}},\ and\
  \bibinfo {author} {\bibfnamefont {S.~R.}\ \bibnamefont {Shenoy}},\ }\href
  {https://doi.org/https://doi.org/10.1103/PhysRevLett.79.4950} {\bibfield
  {journal} {\bibinfo  {journal} {Phys. Rev. Lett.}\ }\textbf {\bibinfo
  {volume} {79}},\ \bibinfo {pages} {4950} (\bibinfo {year}
  {1997})}\BibitemShut {NoStop}%
\bibitem [{\citenamefont {Paraoanu}\ \emph {et~al.}(2001)\citenamefont
  {Paraoanu}, \citenamefont {Kohler}, \citenamefont {Sols},\ and\ \citenamefont
  {Leggett}}]{Paraoanu2001}%
  \BibitemOpen
  \bibfield  {author} {\bibinfo {author} {\bibfnamefont {G.-S.}\ \bibnamefont
  {Paraoanu}}, \bibinfo {author} {\bibfnamefont {S.}~\bibnamefont {Kohler}},
  \bibinfo {author} {\bibfnamefont {F.}~\bibnamefont {Sols}},\ and\ \bibinfo
  {author} {\bibfnamefont {A.~J.}\ \bibnamefont {Leggett}},\ }\href
  {https://doi.org/10.1088/0953-4075/34/23/313} {\bibfield  {journal} {\bibinfo
   {journal} {J. Phys. B}\ }\textbf {\bibinfo {volume} {34}},\ \bibinfo {pages}
  {4689} (\bibinfo {year} {2001})}\BibitemShut {NoStop}%
\bibitem [{\citenamefont {Meier}\ and\ \citenamefont
  {Zwerger}(2001)}]{Meier2001}%
  \BibitemOpen
  \bibfield  {author} {\bibinfo {author} {\bibfnamefont {F.}~\bibnamefont
  {Meier}}\ and\ \bibinfo {author} {\bibfnamefont {W.}~\bibnamefont
  {Zwerger}},\ }\href {https://doi.org/10.1103/PhysRevA.64.033610} {\bibfield
  {journal} {\bibinfo  {journal} {Phys. Rev. A}\ }\textbf {\bibinfo {volume}
  {64}},\ \bibinfo {pages} {033610} (\bibinfo {year} {2001})}\BibitemShut
  {NoStop}%
\bibitem [{\citenamefont {Gati}\ and\ \citenamefont
  {Oberthaler}(2007)}]{Gati2007}%
  \BibitemOpen
  \bibfield  {author} {\bibinfo {author} {\bibfnamefont {R.}~\bibnamefont
  {Gati}}\ and\ \bibinfo {author} {\bibfnamefont {M.~K.}\ \bibnamefont
  {Oberthaler}},\ }\href {https://doi.org/10.1088/0953-4075/40/10/R01}
  {\bibfield  {journal} {\bibinfo  {journal} {J. Phys. B}\ }\textbf {\bibinfo
  {volume} {40}},\ \bibinfo {pages} {R61} (\bibinfo {year} {2007})}\BibitemShut
  {NoStop}%
\bibitem [{\citenamefont {Spuntarelli}\ \emph {et~al.}(2007)\citenamefont
  {Spuntarelli}, \citenamefont {Pieri},\ and\ \citenamefont
  {Strinati}}]{Spuntarelli2007}%
  \BibitemOpen
  \bibfield  {author} {\bibinfo {author} {\bibfnamefont {A.}~\bibnamefont
  {Spuntarelli}}, \bibinfo {author} {\bibfnamefont {P.}~\bibnamefont {Pieri}},\
  and\ \bibinfo {author} {\bibfnamefont {G.~C.}\ \bibnamefont {Strinati}},\
  }\href {https://doi.org/10.1103/PhysRevLett.99.040401} {\bibfield  {journal}
  {\bibinfo  {journal} {Phys. Rev. Lett.}\ }\textbf {\bibinfo {volume} {99}},\
  \bibinfo {pages} {040401} (\bibinfo {year} {2007})}\BibitemShut {NoStop}%
\bibitem [{\citenamefont {Boukobza}\ \emph {et~al.}(2010)\citenamefont
  {Boukobza}, \citenamefont {Moore}, \citenamefont {Cohen},\ and\ \citenamefont
  {Vardi}}]{Boukobza2010}%
  \BibitemOpen
  \bibfield  {author} {\bibinfo {author} {\bibfnamefont {E.}~\bibnamefont
  {Boukobza}}, \bibinfo {author} {\bibfnamefont {M.~G.}\ \bibnamefont {Moore}},
  \bibinfo {author} {\bibfnamefont {D.}~\bibnamefont {Cohen}},\ and\ \bibinfo
  {author} {\bibfnamefont {A.}~\bibnamefont {Vardi}},\ }\href
  {https://doi.org/10.1103/PhysRevLett.104.240402} {\bibfield  {journal}
  {\bibinfo  {journal} {Phys. Rev. Lett.}\ }\textbf {\bibinfo {volume} {104}},\
  \bibinfo {pages} {240402} (\bibinfo {year} {2010})}\BibitemShut {NoStop}%
\bibitem [{\citenamefont {Chen}\ \emph {et~al.}(2020)\citenamefont {Chen},
  \citenamefont {Mukhopadhyay},\ and\ \citenamefont {Schmelcher}}]{chen2020}%
  \BibitemOpen
  \bibfield  {author} {\bibinfo {author} {\bibfnamefont {J.}~\bibnamefont
  {Chen}}, \bibinfo {author} {\bibfnamefont {A.~K.}\ \bibnamefont
  {Mukhopadhyay}},\ and\ \bibinfo {author} {\bibfnamefont {P.}~\bibnamefont
  {Schmelcher}},\ }\href {https://doi.org/10.1103/PhysRevA.102.033302}
  {\bibfield  {journal} {\bibinfo  {journal} {Phys. Rev. A}\ }\textbf {\bibinfo
  {volume} {102}},\ \bibinfo {pages} {033302} (\bibinfo {year}
  {2020})}\BibitemShut {NoStop}%
\bibitem [{\citenamefont {Singh}\ \emph {et~al.}(2016)\citenamefont {Singh},
  \citenamefont {Weimer}, \citenamefont {Morgener}, \citenamefont {Siegl},
  \citenamefont {Hueck}, \citenamefont {Luick}, \citenamefont {Moritz},\ and\
  \citenamefont {Mathey}}]{Singh2016}%
  \BibitemOpen
  \bibfield  {author} {\bibinfo {author} {\bibfnamefont {V.~P.}\ \bibnamefont
  {Singh}}, \bibinfo {author} {\bibfnamefont {W.}~\bibnamefont {Weimer}},
  \bibinfo {author} {\bibfnamefont {K.}~\bibnamefont {Morgener}}, \bibinfo
  {author} {\bibfnamefont {J.}~\bibnamefont {Siegl}}, \bibinfo {author}
  {\bibfnamefont {K.}~\bibnamefont {Hueck}}, \bibinfo {author} {\bibfnamefont
  {N.}~\bibnamefont {Luick}}, \bibinfo {author} {\bibfnamefont
  {H.}~\bibnamefont {Moritz}},\ and\ \bibinfo {author} {\bibfnamefont
  {L.}~\bibnamefont {Mathey}},\ }\href
  {https://doi.org/10.1103/PhysRevA.93.023634} {\bibfield  {journal} {\bibinfo
  {journal} {Phys. Rev. A}\ }\textbf {\bibinfo {volume} {93}},\ \bibinfo
  {pages} {023634} (\bibinfo {year} {2016})}\BibitemShut {NoStop}%
\bibitem [{Note1()}]{Note1}%
  \BibitemOpen
  \bibinfo {note} {$l_x$ is chosen to be smaller than the healing length $\xi =
  \hbar /\protect \sqrt {2 m g_{\protect \mathrm {1D}} n}$ and the thermal de
  Broglie wavelength $\lambda = \protect \sqrt {2\pi \hbar ^2/(m k_B T)}$,
  where $k_B$ is Boltzmann constant and $T$ is the temperature.}\BibitemShut
  {Stop}%
\bibitem [{\citenamefont {Tinkham}(2004)}]{Tinkham2004}%
  \BibitemOpen
  \bibfield  {author} {\bibinfo {author} {\bibfnamefont {M.}~\bibnamefont
  {Tinkham}},\ }\href {http://www.worldcat.org/isbn/0486435032} {\emph
  {\bibinfo {title} {Introduction to Superconductivity}}},\ \bibinfo {edition}
  {2nd}\ ed.\ (\bibinfo  {publisher} {Dover Publications},\ \bibinfo {year}
  {2004})\BibitemShut {NoStop}%
\end{thebibliography}%

\end{document}